# Analytical description of domain morphology and phase diagrams of ferroelectric nanoparticles

*Anna N. Morozovska [1,2] Yevhen M. Fomichov[3], Anna N. Morozovska [1,2] Yevhen M. Fomichov[3],*

*Petro Maksymovych[4], Yulian M. Vysochanskii[5], and Eugene A. Eliseev[3] [*]*

[1] *Institute of Physics, National Academy of Sciences of Ukraine,*

*46, pr. Nauky, 03028 Kyiv, Ukraine*

[2] *Bogolyubov Institute for Theoretical Physics, National Academy of Sciences of Ukraine,*

*14-b Metrolohichna str. 03680 Kyiv, Ukraine*

[3] *Institute for Problems of Materials Science, National Academy of Sciences of Ukraine,*

*Krjijanovskogo 3, 03142 Kyiv, Ukraine*

[4]*Center for Nanophase Materials Sciences, Oak Ridge National Laboratory, Oak Ridge, TN, 37831, USA*

[5] *Institute of Solid State Physics and Chemistry, Uzhgorod University,*

*88000 Uzhgorod, Ukraine*

## Abstract

Analytical description of domain structure morphology and phase diagrams of ferroelectric nanoparticles is developed in the framework of Landau-Ginzburg-Devonshire approach. To model realistic conditions of incomplete screening of spontaneous polarization at the particle surface, it was considered covered by an ultra-thin layer of screening charge with finite screening length. The phase diagrams, calculated for spherical $Sn_2P_2S_6$ nanoparticles in coordinates "temperature – surface screening length" by finite element modeling, demonstrate the emergence of poly-domain region at the tricritical point and its broadening with increasing the screening length for the particle radius over a critical value. Metastable and stable labyrinthine domain structures exist in $Sn_2P_2S_6$ nanoparticles with radius (8-10) nm and more, similarly to the case of $CuInP_2S_6$ nanoparticles considered previously. We derived simple analytical expressions for the boundaries between paraelectric, single-domain and poly-domain ferroelectric phases, tricritical point and the necessary condition for the appearance of labyrinthine domains, and demonstrated their high accuracy in comparison with finite element modeling results.

Analytical expressions for the dependence of the ferroelectric-paraelectric transition temperature on the particle radius in the single-domain and poly-domain states of the particle were compared with analogous dependences experimentally measured for $SrBi_2Ta_2O_9$ nanoparticles and simulated for $Sn_2P_2S_6$ nanocrystals by Monte Carlo method within the framework of axial next-nearest-neighbours Ising model. The analytical expression for the ferroelectric-paraelectric transition temperature in the poly-domain state quantitatively agrees with experimental and simulated results, and it perfectly reproduces empirical Ishikawa equation at all temperatures, justifying it theoretically. Analytical description shows that phase diagrams and domain morphologies, which are qualitatively similar to the ones calculated in this work, can be expected in other ferroelectric nanoparticles covered by the screening charges, being rather different for the ferroelectrics with the first and second order ferroelectric-paraelectric transitions respectively.

---

[*] corresponding author, e-mail: eugene.a.eliseev@gmail.com

## I. INTRODUCTION

Nanoscale ferroelectrics are unique model objects for fundamental studies of polar surface properties, various screening mechanisms of spontaneous polarization by free carriers and possible emergence of versatile multi-domain states [1, 2], including their complex morphologies [3, 4, 5, 6]. Ferroelectric nanoparticles can demonstrate perfect possibilities of polar-active properties control, which attract permanent attention of researchers. Classical examples are nontrivial experimental results of Yadlovker and Berger [7, 8, 9], which reveal the enhancement of polar properties of cylindrical nanoparticles of Rochelle salt. Frey and Payne [10], Zhao et al [11], Drobnich et al [12], Erdem et al [13], Shen et al [14] and Golovina et al [15, 16, 17] demonstrated the possibility to control by finite size effects the phase transition temperatures and other peculiarities, including the appearance of new polar phases for $BaTiO_3$, $S_2P_2S_6$, $PbTiO_3$, $SrBi_2Ta_2O_9$ and $KTa_{1-x}Nb_xO_3$ nanopowders and nanoceramics, respectively. Increasing interest is related with the impact of surface conditions and finite size effects on the photorefractive properties of $S_2P_2S_6$ or $BaTiO_3$ ferroelectric nanoparticles suspended in nematic liquid crystals (see [18, 19] and refs therein).

The continuum phenomenological Landau-Ginzburg-Devonshire (**LGD**) approach combined with the electrostatic equations allows one to establish the physical origin of the anomalies in the polar and dielectric properties, and calculate the phase diagrams changes appearing under the decrease of ferroelectric particle sizes. For instance, using the **LGD** approach Niepce [20], Huang et al [21, 22], Glinchuk et al [23, 24], Ma [25], Khist et al [26], Wang et al [27], Morozovska et al [28, 29, 30, 31] and Eliseev et al [32, 33, 34] have shown, that the changes of the transition temperatures, enhancement or weakening of spontaneous polar or/and magnetic order in a single-domain spherical, ellipsoidal and cylindrical nanoparticles of sizes (4 – 100 nm) are conditioned by the various physical mechanisms, such as surface tension, correlation effect, depolarization field, flexoelectricity, electrostriction, magnetoelectric coupling, magnetostriction, rotostriction and Vegard-type chemical pressure. We emphasize that the applicability of the LGD approach for ferroelectric nanoparticles with sizes about (4-5) nm or more (i.e. 10 lattice constants or more) is corroborated by the fact, that the critical sizes for the appearance of the long-range order and the properties calculated for thin films or nanoparticles from atomistic [35 ,36, 37, 38, 39] and phenomenological [32-30, 40, 41] theories are in a good agreement with each other as well as with experimental results for nanosized ferromagnetics [42] and ferroelectrics [7-11, 13, 43]. As a mean-field approach, LGD loses its validity below 5-10 unit-cells due to the vanishing of long-range order correlations.

Incomplete screening of spontaneous polarization causes depolarization fields, which in turn can lead to appearance of ferroelectric domains in the particle, decreasing the positive energy of depolarization field [26, 27, 33, 34, 44]. Incomplete screening conditions of the spontaneous polarization also leads to the decrease of ferroelectric transition temperature due to the depolarization field effect [26, 27, 33, 34]. Yet the vast majority of theoretical models (both LGD-based and *ab initio*) consider the particles covered with perfect electrodes, stabilizing their single-domain state (see e.g. [20-32]). Only several LGD-based theoretical studies did consider the incomplete screening of spontaneous polarization in ferroelectric nanoparticles [26, 27, 33, 34, 45]. In particular, Eliseev et al [45] calculated the phase diagram and domain structure morphology in spherical nanoparticles of uniaxial ferroelectric $CuInP_2S_6$ covered by a layer of screening charge with finite screening length. They revealed that a regular stripe domain structure transforms into a labyrinth pattern when the gradient term decreases below the critical value, and classified the event as a gradient-induced morphological transition

Under incomplete screening conditions, the analytical description of the domain structure morphology changes and phase diagrams of ferroelectric nanoparticles is absent. In particular, any sort of analytical expressions for the transition temperatures between different poly-domain, single-domain and paraelectric phases is absent. The available analytical expressions for the transition between single-domain and paraelectric phases give essentially underestimated values of the particle critical sizes [33]. To fill the knowledge gap, here we propose LGD-based analytical description of domain structure morphology and phase diagrams of ferroelectric nanoparticles. To model realistic conditions of spontaneous polarization incomplete screening at the particle surface, it was regarded covered by an ultra-thin layer of screening charge with finite screening length

The manuscript is structured as follows. Free energy and basic equations with boundary conditions are discussed in **section II**. Analytical expressions for the phase boundaries separating single-domain, poly-domain and paraelectric phases are listed and analyzed in **section III**. Phase diagrams of ferroelectric nanoparticles covered by sluggish screening charges are analyzed in **section IV** with the special attention to domain morphologies changes and temporal evolution of striped and labyrinthine domains**.** Comparison with available experiments [14] for $SrBi_2Ta_2O_9$ nanoparticles and independent Monte Carlo (**MC**) simulations within axial next-nearest-neighbours Ising (**ANNNI**) model [12] for $S_2P_2S_6$ nanocrystalls is presented in **section V. Section VI** is a brief discussion with conclusive remarks. Electrostatic problem, derivation of analytical expressions and phase diagrams calculated for nanoparticles of different radius are given in **Appendixes A, B** and **C**, respectively.

## II. PROBLEM STATEMENT

Let us consider a ferroelectric nanoparticle of radius $R$ with a one-component ferroelectric polarization $P_3(\mathbf{r})$ directed along the crystallographic axis 3 [**Fig.1(a)**]. At the same time we can assume that the dependence of other electric polarization components on the inner field electric $E_i$ is linear $P_i = \varepsilon_0(\varepsilon_b - 1)E_i$, where $i = 1, 2$ and $\varepsilon_b$ is an isotropic relative permittivity of background [46]. Since the ferroelectric polarization component $P_3(\mathbf{r})$ contains background and soft mode contributions, electric displacement vector has the form $\mathbf{D} = \varepsilon_0\varepsilon_b\mathbf{E} + \mathbf{P}$ inside the particle. Outside the partilce $\mathbf{D} = \varepsilon_0\varepsilon_e\mathbf{E}$, where $\varepsilon_e \sim 1$ is the relative dielectric permittivity of external media (air or vacuum).

To model realistic conditions of spontaneous polarization incomplete screening at the particle surface, we regard that the particle surface is covered by an ultra-thin layer of "sluggish" screening charge with a surface charge density σ, at that σ linearly depends on electric potential φ at the surface, $\sigma \approx -\varepsilon_0 \varphi/\Lambda$, where $\varepsilon_0$ is a universal dielectric constant and $\Lambda$ is a surface screening length [26, 33, 34, 45]. In many cases the screening charges can be localized at Bardeen-type surface states [47] caused by the strong band-bending via depolarization field [48, 49, 50, 51, 52], and for the case $\Lambda$ can be much smaller (≤1 Å) than a lattice constant (~0.5 nm) [27] and almost temperature-independent. Another important case (relevant to the nanoparticles suspension in liquid crystals) is the Stephenson-Highland (**SH**) ionic adsorption at the ferroelectric surface [53, 54]. The linearization of Langmuir absorption isotherm [55] leads to the expression for $\Lambda^{-1} \approx \sum_i \frac{(eZ_i)^2}{4\varepsilon_0 A_i k_B T}\left(1 - \tanh^2\left(\frac{\Delta G_i^{00}}{2k_B T}\right)\right)$, where $Z_i$ is the ionization number of the surface ions, $T$ is the absolute temperature, $1/A_i$ are their saturation densities of positive and negative ionic species ($i$=1,2), $\Delta G_i^{00}$ are the free energies of the surface ions formation in SH model. Since we would not like to probe a range of temperature dependence of $\Lambda$, we performed calculations regarding $\Lambda$ changing in the range ($10^{-3}$ – 1) Å.

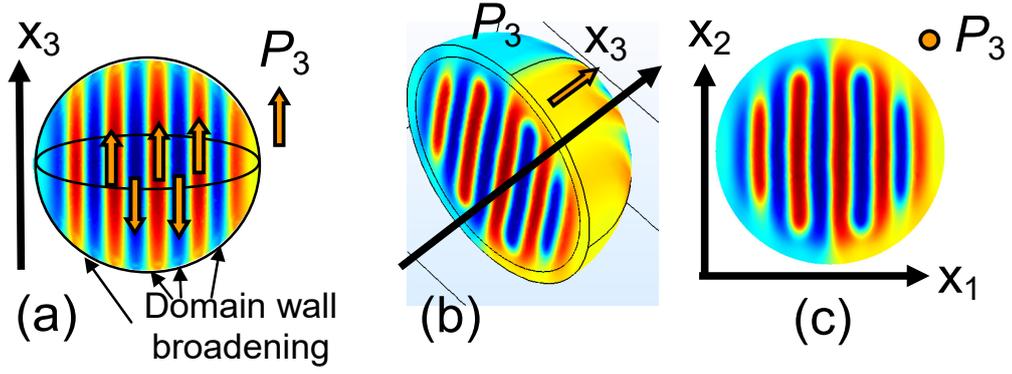

**FIG. 1.** Domain stripes calculated in a spherical $Sn_2P_2S_6$ nanoparticle of radius $R=10$ nm, screening length $\Lambda=0.15$ Å and room temperature. **(a)** Polar cross-section, **(b)** tilted semi-spherical view and **(c)** equatorial cross-section.

LGD free energy functional $G$ additively includes 2-4-6 Landau expansion on polarization powers, $G_{Landau}$, polarization gradient energy contribution, $G_{grad}$, electrostatic contribution $G_{el}$, and elastic, electrostriction and flexoelectric contributions $G_{es+flexo}$:

$$G = G_{Landau} + G_{grad} + G_{el} + G_{es+flexo}, \tag{1a}$$

$$G_{Landau} = \int_{|\vec{r}|<R} d^3r \left( \frac{\alpha}{2} P_3^2 + \frac{\beta}{4} P_3^4 + \frac{\gamma}{6} P_3^6 \right), \tag{1b}$$

$$G_{grad} = \int_{|\vec{r}|<R} d^3r \left( \frac{g_{11}}{2} \left(\frac{\partial P_3}{\partial x_3}\right)^2 + \frac{g_{44}}{2} \left[ \left(\frac{\partial P_3}{\partial x_2}\right)^2 + \left(\frac{\partial P_3}{\partial x_1}\right)^2 \right] \right), \tag{1c}$$

$$G_{el} = -\int_{|\vec{r}|<R} d^3r \left( P_3 E_3 + \frac{\varepsilon_0 \varepsilon_b}{2} E_i E_i \right) - \int_{|\vec{r}|=R} d^2r \left( \frac{\varepsilon_0 \varphi^2}{2\Lambda} \right) - \frac{\varepsilon_0 \varepsilon_e}{2} \int_{|\vec{r}|>R} E_i E_i d^3r, \tag{1d}$$

$$G_{es+flexo} = \int_{|\vec{r}|<R} d^3r \left( -\frac{s_{ijkl}}{2} \sigma_{ij} \sigma_{kl} - Q_{ij3} \sigma_{ij} P_3^2 - F_{ijk3} \left( \sigma_{ij} \frac{\partial P_3}{\partial x_k} - P_3 \frac{\partial \sigma_{ij}}{\partial x_k} \right) \right). \tag{1e}$$

The coefficient $\alpha$ linearly depends on temperature $T$, $\alpha = \alpha_T (T - T_C)$, where $T_C$ is the Curie temperature and $\alpha_T$ is the inverse Curie-Weiss constant. The coefficient $\beta$ is regarded temperature-independent. It is positive if the ferroelectric material undergoes a second order transition to the paraelectric phase at $T_C$ and negative otherwise. Higher nonlinear coefficient $\gamma$ and gradient coefficients $g_{11}$ and $g_{44}$ are positive and regarded temperature independent. $\sigma_{ij}$ is the stress tensor in Eq.(1e). Electric field components $E_i$ are related with electric potential $\varphi$ as $E_i = -\partial \varphi / \partial x_i$

We omit the explicit form of the $G_{es+flexo}$ for the sake of simplicity, it is listed in Refs.[56, 57, 58]. Since the values of the electrostriction and flexoelectric tensor components, $Q_{ijkl}$ and $F_{ijkl}$, are unknown for most of ferroelectrics, we performed numerical calculations with the coefficients varied in a physically reasonable range ($|F_{ijkl}| \leq 10^{11}$ m$^3$/C and $|Q_{ijkl}| \leq 0.1$ m$^4$/C$^2$) and obtained results proved the insignificant impact of electrostriction and flexoelectric coupling on domain morphology for most of ferroics [45].

Minimization of the free energy (1) with respect to polarization $P(\mathbf{r}_3)$ leads to the Euler-Lagrange equation for its value

$$\alpha(T)P_3 + \beta P_3^3 + \gamma P_3^5 - g_{44}\left(\frac{\partial^2}{\partial x_1^2} + \frac{\partial^2}{\partial x_2^2}\right)P_3 - g_{11}\frac{\partial^2 P_3}{\partial x_3^2} = E_3. \qquad (2)$$

The boundary condition for polarization at the spherical surface $r=R$ is natural, $\partial \vec{P}_3/\partial \mathbf{n}\big|_{r=R} = 0$, $\mathbf{n}$ is the outer normal to the surface. The potential $\varphi$ satisfies a Poisson equation inside the particle and Laplace equation outside it,

$$\varepsilon_0 \varepsilon_b \left(\frac{\partial^2}{\partial x_1^2} + \frac{\partial^2}{\partial x_2^2} + \frac{\partial^2}{\partial x_3^2}\right)\varphi = \begin{cases} -\dfrac{\partial P_3}{\partial x_3}, & r < R, \\ 0, & r > R \end{cases}, \qquad (3)$$

Equations (3) are supplemented by the condition of potential continuity at the particle surface, $(\varphi_{ext} - \varphi_{int})\big|_{r=R} = 0$. The boundary condition for the normal components of electric displacements, $(\mathbf{n}(\mathbf{D}_{ext} - \mathbf{D}_{int}) + \sigma)\big|_{r=R} = 0$, where the surface charge density $\sigma = -\varepsilon_0 \varphi/\Lambda$ is linearly proportional to the electric potential and inversely proportional to the screening length.

### III. ANALYTICAL EXPRESSIONS FOR THE PHASE BOUNDARIES

Phase diagrams of spherical ferroelectric nanoparticles covered by a screening charge have several phases, namely paraelectric (**PE**) phase, single-domain ferroelectric (**SDFE**) phase and poly-domain ferroelectric (**PDFE**) phase including various domain morphologies [45]. Our target is to derive present accurate approximate analytical expressions for the phase boundaries and compare the analytical formulas with finite element modeling (**FEM**) modeling.

Approximate expression for the nanoparticle transition temperature from SDFE to PE phase is

$$T_{PE-SDFE}(R, \Lambda) = T_C - \frac{1}{\alpha_T \varepsilon_0 [\varepsilon_b + 2\varepsilon_e + (R/\Lambda)]} \qquad (4)$$

Here the first term is a bulk Curie temperature. The second term is originated from depolarization field. The term, depending on the ratio $(R/\Lambda)$, strongly decreases the PE-SDFE transition temperature of small nanoparticles and vanishes for big particles being proportional to $(\Lambda/R)$. Derivation of Eq.(4) is given in Ref. [33] and also listed in **Appendix A.** Notably that the expression (4) is exact for the natural boundary conditions at the particle surface, $\left.\partial \vec{P}/\partial \mathbf{n}\right|_{r=R} = 0$, and becomes high-accuracy approximation for more general boundary condition, when $\left.\vec{P} + \lambda \partial \vec{P}/\partial \mathbf{n}\right|_{r=R} = 0$, where the so-called extrapolation length λ [59, 60] is regarded positive and typically exceeds (1 – 2) nm [61]. Hence Eq.(4) can be used to check the FEM simulations accuracy and convergence rate for the natural boundary conditions used hereinafter.

Approximate analytical expression for the nanoparticle transition temperature from PDFE to PE phase can be found using calculations in the **Appendix B**, namely:

$$T_{PE-PDFE}(R,\Lambda) = \begin{cases} T_C - \dfrac{g_{44}}{\alpha_T \xi R}\left(\dfrac{1}{\sqrt{\varepsilon_0(\varepsilon_b + 2\varepsilon_e)g_{44}}} + \dfrac{1}{\xi R_{cr}(\Lambda)} - \dfrac{1}{\xi R}\right), & R > R_{cr} \\ \text{absent}, & R < R_{cr} \end{cases} \quad (5a)$$

Where the critical radius is

$$R_{cr}(\Lambda) = \left(\dfrac{\xi}{\sqrt{\varepsilon_0(\varepsilon_b + 2\varepsilon_e)g_{44}}} - \dfrac{1}{(\varepsilon_b + 2\varepsilon_e)\Lambda}\right)^{-1}. \quad (5b)$$

Three terms in brackets in Eq.(5a) originated from the correlation effect and depolarization field energy of the domain stripes. Parameter $\xi$ is a sort of geometrical factor that is close to 0.5 for domain stripes onset in the {x,y} cross-section of the spherical particle.

The minimal spatial wave number $k_{min} = \sqrt{k_x^2 + k_y^2}$ and period $D_{max}$ of the domain structure onset in the {x,y} cross-section are radius-dependent and temperature-independent,

$$k_{min}(R,\Lambda) = \dfrac{1}{\xi R}\sqrt{\dfrac{R}{R_{cr}(\Lambda)} - 1}, \qquad D_{max} = \dfrac{2\pi}{k_{min}}. \quad (6)$$

Notably, that the identical form of Eq.(5) can be rewritten via $k_{min}$ as

$$T_{PE-PDFE} \approx T_C - \dfrac{1}{\alpha_T}\left(g_{44}k_{min}^2 + \dfrac{\varepsilon_0^{-1}}{(R/\Lambda) + (\varepsilon_b + 2\varepsilon_e)(1 + (\xi R)^2 k_{min}^2)}\right) \quad \text{(\textbf{Appendix B} for details).}$$

Expressions (5)-(6) are valid under the condition

$$\frac{\xi}{\sqrt{\varepsilon_0(\varepsilon_b + 2\varepsilon_e)g_{44}}} \geq \frac{1}{R} + \frac{1}{(\varepsilon_b + 2\varepsilon_e)\Lambda}. \tag{7}$$

Equation (7) means that the critical value of the gradient coefficient exists at fixed other parameters, $g_{44}^{cr}(R,\Lambda) = \frac{\xi^2}{\varepsilon_0(\varepsilon_b + 2\varepsilon_e)}\left(\frac{1}{R} + \frac{1}{(\varepsilon_b + 2\varepsilon_e)\Lambda}\right)^{-2}$, and domains appears at $g_{44} < g_{44}^{cr}(R,\Lambda)$. At fixed gradient coefficient $g_{44}$ the equality in Eq.(7) means that the relation between the particle radius $R$ and screening length $\Lambda$ should be valid for the domain onset.

The fulfillment of the equality in Eq.(7) corresponds to the transition to a single domain state that occurs in a three-critical point on the phase diagram, where the energies of SDFE and PDFE phases are equal to zero energy of PE phase. In the three-critical point $k_{min} = 0$ and $T_{PE-PDFE}(R,\Lambda) = T_{PE-SDFE}(R,\Lambda)$ allowing for Eqs.(5)-(6). After substitution into Eq.(4), the radius dependence of the tricritical point temperature $T_{tcr}(R)$ and screening length $\Lambda_{tcr}(R)$ can be found exactly as

$$T_{tcr}(R) = T_C - \frac{\sqrt{\varepsilon_0(\varepsilon_b + 2\varepsilon_e)g_{44}}}{\alpha_T \varepsilon_0(\varepsilon_b + 2\varepsilon_e)\xi R}, \tag{8a}$$

$$\frac{1}{\Lambda_{tcr}(R)} = (\varepsilon_b + 2\varepsilon_e)\left(\frac{\xi}{\sqrt{\varepsilon_0(\varepsilon_b + 2\varepsilon_e)g_{44}}} - \frac{1}{R}\right). \tag{8b}$$

From these expression the radius dependences $T_{tcr}(R)$ and $\Lambda_{tcr}^{-1}(R)$ scales as $1/R$.

Approximate analytical expression for the nanoparticle transition temperature from PDFE to SDFE phase can be estimated from the free energy equality of the phases, since the transition is of the first order. Using the speculations we derived the expansion

$$T_{PD-SD}(R,\Lambda) \approx T_{tcr}(R) - \Delta T(R)\left(1 - \frac{\Lambda}{\Lambda_{tcr}(R)}\right)^\delta. \tag{9}$$

Power $\delta$ is related with the critical index of the phase transition and hence should be radius-independent. The radius dependence of the temperature shift $\Delta T(R)$ should be established from FEM. Notably Eqs.(9) is valid only when inequality in Eq.(7) is fulfilled.

### IV. PHASE DIAGRAMS FOR Sn$_2$P$_2$S$_6$ NANOPARTICLES

LGD parameters for a bulk ferroelectric Sn$_2$P$_2$S$_6$ (**SPS**) were collected from Refs.[62, 63] and references therein. They are listed in **Table I**.

Table I. LGD parameters for bulk ferroelectric $Sn_2P_2S_6$

| $\varepsilon_b$ | $\alpha_T(C^{-2} \cdot m\, J/K)$ | $T_C$ (K) | $\beta$ ($C^{-4} \cdot m^5 J$) | $\gamma$ ($C^{-6} \cdot m^9 J$) | $g_{11}$ ($m^3/F$) | $g_{44}$ ($m^3/F$) |
|---|---|---|---|---|---|---|
| 7 | $1.6 \times 10^6$ | 336 | $7.42 \times 10^8$ | $3.5 \times 10^{10}$ | $3.0 \times 10^{-10}$ | $(0.1 - 1) \times 10^{-10}$ |

In **Appendix C** we presented the phase diagrams of SPS nanoparticles with radius $R =(2 – 10)$ nm calculated in coordinates "temperature $T$ – screening length $\Lambda$" for the gradient coefficients $g_{44}=10^{-11}$ m$^3$/F and $g_{44}=10^{-10}$ m$^3$/F, which are one order of magnitude different. The particles with radius less than 4 nm are either paraelectric or single-domain for all $\Lambda$ values changing in the range $(10^{-3} – 1)$ Å [see **Fig.S1** in **Appendix C**]. Different domain morphologies appear for 4-nm and bigger particles with $\Lambda$ increase of more than 0.1 Å [see **Fig.S2-S3** in **Appendix C**].

A typical phase diagrams of 10-nm SPS nanoparticles calculated in coordinates "temperature $T$ – screening length $\Lambda$" are shown in **Fig. 2**. **Fig. 2(a)** corresponds to the gradient coefficient $g_{44}=10^{-11}$ m$^3$/F and in **Fig. 2(b)** is for $g_{44}=10^{-10}$ m$^3$/F. At small $g_{44}$ the phase diagram has much wider region of stable poly-domain states (**PDFE**) separating the single-domain ferroelectric (**SDFE**) and nonpolar paraelectric (**PE**) phases (compare **Fig.2(a)** and **2(b)**). Abbreviation "*trc*" denotes the tricritical point at the diagram. The tricritical point has coordinates $\{\Lambda_{tcr}, T_{tcr}\}$.

Labyrinthine domain (**LD**) region exists for small gradient coefficient $g_{44} \leq 10^{-11}$ m$^3$/F, and is absent for its higher values, $g_{44} \geq 10^{-10}$ m$^3$/F. Since LDs are stable over all computation time, they are either absolutely stable or at least metastable. However absolute stability of LDs is questionable, because bulk SPS undergoes the second order phase transition at $T_C$, while it was expected that only ferroelectric nanoparticles with the first order phase transitions in the bulk can exhibit absolutely stable LD [45].

The bottom row **(I)** in **Fig. 2** shows the typical changes of polarization distribution in the equatorial cross-section of the 10-nm nanoparticle, which happens with $\Lambda$ increase. A SDFE state is stable at very small $\Lambda<0.1$ Å**,** two-domain and poly-domain states are stable in the interval 0.11 Å $<\Lambda<0.14$ Å. Coexistence of PDFE with mixture of domain stripes, LD and PE state appears at 0.145 Å $<\Lambda<2$ Å and is followed by the size-induced phase transition into a stable PE at $\Lambda>10$ Å.

The bottom row **(II)** in **Fig. 2** illustrates the polarization distributions in the stable (**SLD**) and metastable (**MLD**) labyrinthine domains calculated with temperature $T$ increase from 230 K to 260 K in the equatorial cross-sections of the nanoparticles with radius $R$=10 nm, $g_{44}=10^{-11}$ m$^3$/F and $\Lambda= 0.3$Å. LDs are stable at $T<240$K, becomes metastable at higher temperatures (240 K$<T<260$ K) and then transforms into a PE phase at T ≥ 260 K. In the temperature range 240 K<T<260 K, where LDs are metastable, the stable are domain stripes coexisting with PE phase at the particle boundary. Notably that the metastabilty or stability of LDs and stripes were concluded from the comparison of the free energies corresponding to these domain morphologies.

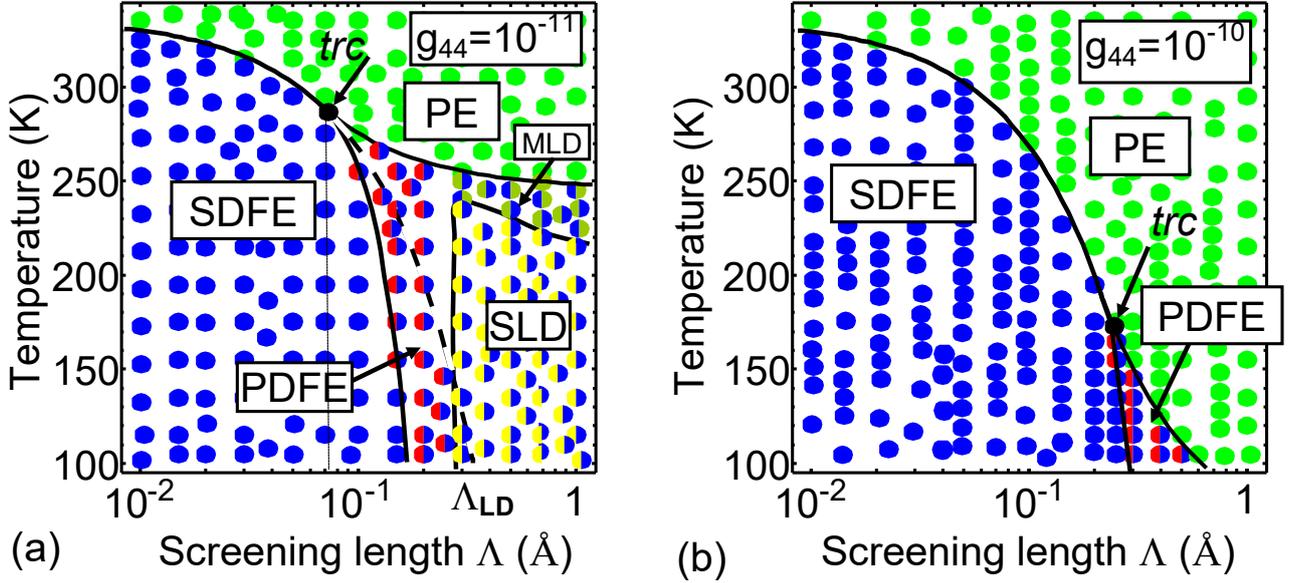

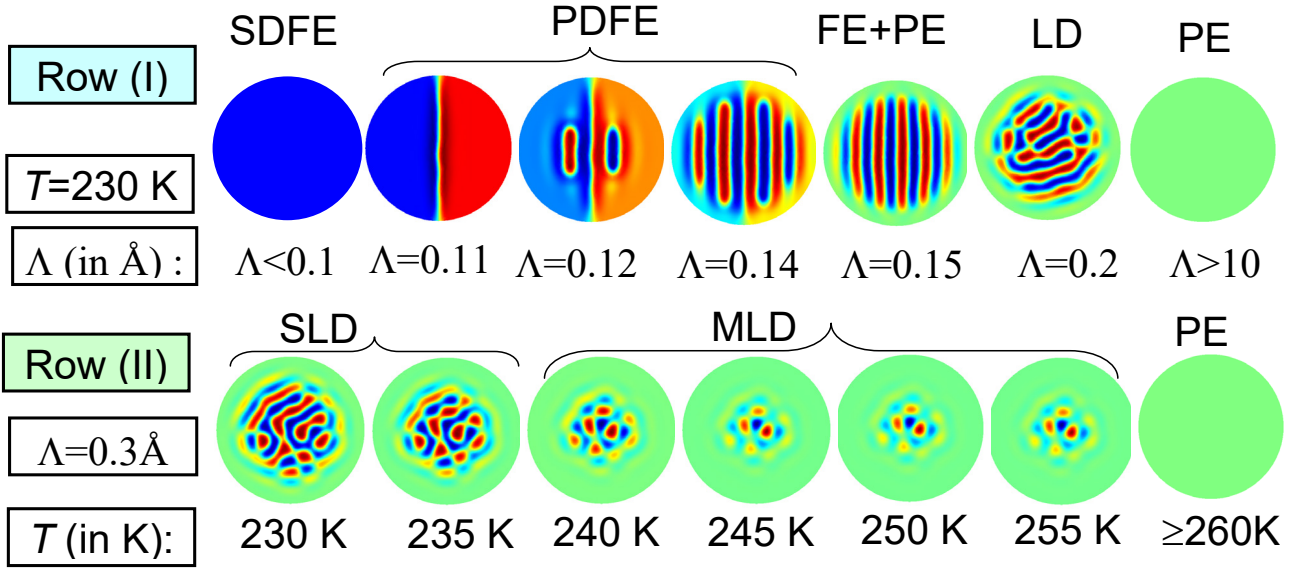

**FIG. 2.** Phase diagram of SPS nanoparticles in coordinates "temperature $T$ – screening length $\Lambda$" calculated for the particle radius 10 nm, gradient coefficient $g_{44}=10^{-11}$ m$^3$/F **(a)** and $g_{44}=10^{-10}$ m$^3$/F **(b)**, ambient permittivity $\varepsilon_e = 1$. The ferroelectric single domain (SDFE), ferroelectric poly domain (PDFE), labyrinthine domain (LD) and paraelectric (PE) phases are stable. Abbreviation "trc" denotes the tricritical point with coordinates $\{\Lambda_{tcr}, T_{tcr}\}$. Solid curves corresponding to the SDFE-PE, PDFE-PE and SDFE-PDFE phase boundaries are calculated from Eqs.(4), (5) and (9), respectively. The bottom row **(I)** shows typical polarization distributions in the equatorial cross-sections of the nanoparticles with radius $R=10$ nm and different values of $\Lambda$ (in Å) for plot

(a). The bottom row **(II)** illustrates the polarization distributions in the stable (SLD) and metastable (MLD) labyrinthine domains calculated with temperature increase from 230 K to 260 K in the equatorial cross-sections of the nanoparticles with radius $R$=10 nm, $g_{44}$=10$^{-11}$ m$^3$/F and $\Lambda$= 0.3Å. SPS parameters are listed in **Table I.**

The temporal evolution of LDs in a 10-nm SPS nanoparticle, calculated using Landau-Khalatnikov relaxation in time-dependent LGD equations [2], is shown by the top line in **Fig. 3.** LDs growth up from a random distribution of small polarization [see **Fig.3(a)**] that's periphery at first becomes paraelectric with time [see **Figs.3(b)-(e)**], and only then the random domains in the central part start growing and transform into the labyrinth with computation time increase [see **Figs.3(f)-(h)**]. At the same other conditions multiple domain stripes can occur from the two-domain configuration shown in **Fig.3(i)**. The stripes evolution is shown in the bottom line in **Figs. 3(j)-(p).**

Notably that the energy values computed for the metastable multiple domain stripes [shown in **Fig.3(p)**] and the stable labyrinthine domain structure [shown in **Fig.3(h)**] are rather different, $G = -4.18 \times 10^{-19}$ J for stripes and $G = -7.62 \times 10^{-19}$ J for LDs. Thus the stable labyrinthine structure has essentially smaller energy than domain stripes corresponding the optimal balance between the gradient-correlation energy (1c) tending to minimize the area of the domain walls (and hence to decrease the number of them) and electrostatic energy (1d) decreasing with reducing domain width. Note that the walls of LDs are uncharged in the central part of the particle and become charged and broadened near its poles [see light regions near the poles in **Figs. 1(a)**], since their broadening causes the decrease of depolarization field [64].

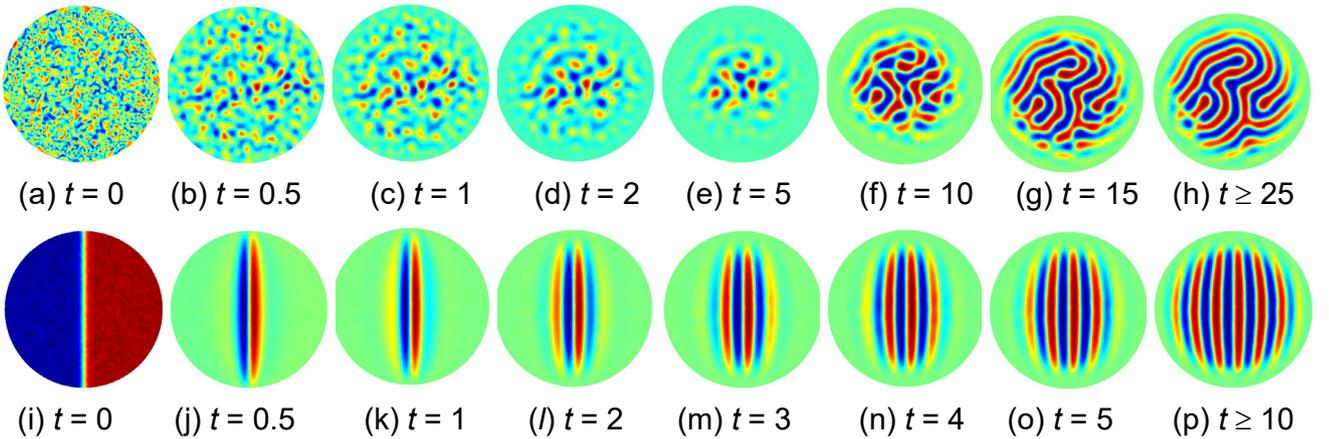

FIG. 3. Temporal evolution of polarization distribution in the equatorial cross-section of the SPS nanoparticle with $R$=10 nm, $\Lambda$=0.1Å, $g_{44}$=2×10$^{-11}$ m$^3$/F and T=200 K calculated starting from random seeding [plots **(a)** – **(h)**] and starting from the two domain configuration [plots **(i)** – **(p)**]. Several moments of dimensionless time $t$ (in the units of Landau-Khalatnikov relaxation time) are shown. SPS parameters are listed in **Table I.**

We leave for further studies the question how the ranges of LDs stability and metastability at phase diagram can be derived analytically (i.e. the boundary between SLD and MLD regions is guided by eye). However the necessary conditions of LDs appearance were derived analytically with the numerical factor estimated from FEM results allowing for the fact that the wave number $k_{min} = \sqrt{k_x^2 + k_y^2}$ given by Eq.(6) is not limited to the description of domain stripes with e.g. $k_x^2 = 0$ and $k_{min} = |k_y|$. Since LDs onset presents an instability in both x- and y-directions (somehow similar to the "chess structure" of its Fourier principal components), corresponding instability condition can be described by Eqs.(7), but with geometrical factor different from ξ. Hence the necessary conditions of LD appearance, which impose inequalities on the radius $R$, lengths $\Lambda$ and gradient coefficient $g_{44}$, have the form

$$\frac{1}{\Lambda} \leq \frac{\eta\sqrt{\varepsilon_b + 2\varepsilon_e}}{\sqrt{\varepsilon_0 g_{44}}} - \frac{\varepsilon_b + 2\varepsilon_e}{R} \quad \text{at fixed } R \text{ and } g_{44}, \quad (10a)$$

$$\frac{1}{R} \leq \frac{\eta}{\sqrt{\varepsilon_0(\varepsilon_b + 2\varepsilon_e)g_{44}}} - \frac{1}{(\varepsilon_b + 2\varepsilon_e)\Lambda} \quad \text{at fixed } \Lambda \text{ and } g_{44}, \quad (10b)$$

$$g_{44} \leq \frac{\eta^2}{\varepsilon_0(\varepsilon_b + 2\varepsilon_e)} \left( \frac{1}{R} + \frac{1}{(\varepsilon_b + 2\varepsilon_e)\Lambda} \right)^{-2} \quad \text{at fixed } R \text{ and } \Lambda. \quad (10c)$$

Here the parameter η has the same sense that the geometrical factor ξ in Eqs.(5), but it appeared close to $1/2\pi$. Inequalities (10) are equivalent to the inequality (7) with the substitution $\xi \to \eta$, and so they are also temperature-independent. Notably that equality in Eq.(10a) reached at $\Lambda = \Lambda_{LD} = \frac{\eta\sqrt{\varepsilon_b + 2\varepsilon_e}}{\sqrt{\varepsilon_0 g_{44}}} - \frac{\varepsilon_b + 2\varepsilon_e}{R}$ determines the almost vertical boundary between the stable LD and domain stripes corresponding to normal PDFE in **Fig.2(a)**.

Noteworthy that the equality in Eq.(10c) gives the critical value of gradient coefficient $g_{44}^{LD} = \frac{\eta^2}{\varepsilon_0(\varepsilon_b + 2\varepsilon_e)} \left( \frac{1}{R} + \frac{1}{(\varepsilon_b + 2\varepsilon_e)\Lambda} \right)^{-2}$ for LDs onset. Actually LDs can exist at $g_{44} < g_{44}^{LD}$. The analytical expression for $g_{44}^{cr}$ corroborates the conclusion made in Ref.[45] about the gradient-induced nature of morphological transition from domain stripes to LDs. However, as it follows from Eqs.(10a) and (10b), the conclusion [45] is incomplete, primary because the inequalities relate all three values, $R$, $\Lambda$ and $g_{44}$. Obtained analytical results lead to the conclusion that more correct statement is that the nature of morphological transition from domain stripes to LDs in ferroelectric nanoparticles underlies in the interplay between the 2D instability of domain splitting induced by incomplete surface screening facilitated by small enough gradient energy and high enough particle size to prevent the effect of geometric catastrophe. Quantitative criteria of LDs appearance based on Eqs.(10) has the form

$$g_{44}\left(\frac{1}{R} + \frac{1}{(\varepsilon_b + 2\varepsilon_e)\Lambda}\right)^2 \leq \frac{\eta^2}{\varepsilon_0(\varepsilon_b + 2\varepsilon_e)}.$$ Since $\frac{1}{R} \ll \frac{1}{(\varepsilon_b + 2\varepsilon_e)\Lambda}$ for typical values $R \geq 4$ nm, $\Lambda$<1Å and $\eta \approx 1/2\pi$, the latter inequality reduces to $\frac{L_d}{\Lambda} \leq \frac{\sqrt{\varepsilon_b + 2\varepsilon_e}}{2\pi}$, where $L_d = \sqrt{\varepsilon_0 g_{44}}$ is the depolarization length. The physical sense of the condition is that the ratio of depolarization to screening lengths should be smaller than effective "geometrical factor" $2\pi\sqrt{\varepsilon_b + 2\varepsilon_e}$.

Analytical expressions for the PE-SDFE, PE-PDFE and SDFE-PDFE boundaries at the phase diagrams have been listed in the **section III** and below we demonstrate their accuracy in comparison with numerical results.

Solid curves in **Fig.2(a)** corresponding to the SDFE-PE, PDFE-PE and SDFE-PDFE phase boundaries are calculated from Eqs.(4), (5) and (9) respectively. The simple analytical expression (4) for the PE-SDFE transition temperature is exact for the natural boundary conditions at the particle surface, and this is the case illustrated in **Fig.2** and supplementary **Figs.S1-S3** in **Appendix C.** The relatively simple analytical expressions (5) for the PE-PDFE transition temperature have very high accuracy and corresponding curves look almost exact in comparison with the PE-PDFE boundary simulated numerically. Expressions (5) contain only one fitting parameter $\xi$ originated from the spherical geometry of the particle. We additionally checked that the fitting value $\xi \approx 0.5$ is not material specific, because it is the same for $SrBi_2Ta_2O_9$, $CuInP_2S_6$ or $LiNbO_3$ nanoparticles. So that Eq.(5) can be used for description of PE-PDFE transition in other nanoparticles of uniaxial ferroelectrics. Analytical expressions (8) for the radius dependence of the critical point temperature $T_{tcr}(R)$ and screening length $\Lambda_{tcr}(R)$ are almost exact. Analytical expression (9) quantitatively describes the boundary between SDFE and PDFE phases with two fitting parameters, temperature change $\Delta T(R)$ and critical index $\delta$, which depend on the material parameters, and therefore this expression is not universal. Parameters $\xi$, $\Delta T(R)$ and power $\delta$ have been extracted from FEM results shown in **Figs.S1-S3** in **Appendix C.** It turned out that the critical index $\delta = 3/2$.

Radius dependences of the dimensionless geometrical factor $\xi$ and temperature $\Delta T(R)$ are shown in **Figs.4(a)-(b)**, respectively. As one can see from **Fig. 4(a)** the geometrical factor $\xi$ slightly increases from 0.44 to 0.49 with the particle radius $R$ increase from 4 nm to 10 nm and then quickly enough saturates to the value around 0.5 with further increase of $R$. The slight increase and fast saturation of $\xi(R)$ mean that this fitting parameter is chosen enough successfully. Additional calculations proved that the value $\xi(R) \approx 0.5$ can be used for description of the size effects in other spherical nanoparticles of uniaxial ferroelectrics such as for $SrBi_2Ta_2O_9$, $CuInP_2S_6$ or $LiNbO_3$.

As one can see from **Fig. 4(b)** the temperature $\Delta T(R)$ rapidly decreases from 105 K to 50 K with increasing the particle radius $R$ from 4 nm to 10 nm and then continues to decrease with further increase of $R$. The rapid decrease of $\Delta T(R)$ and the absence of its saturation $R$ increase mean that the choice of this fitting parameter is not very successful. However its noticeable decrease with $R$ increase appeared universal for other spherical nanoparticles.

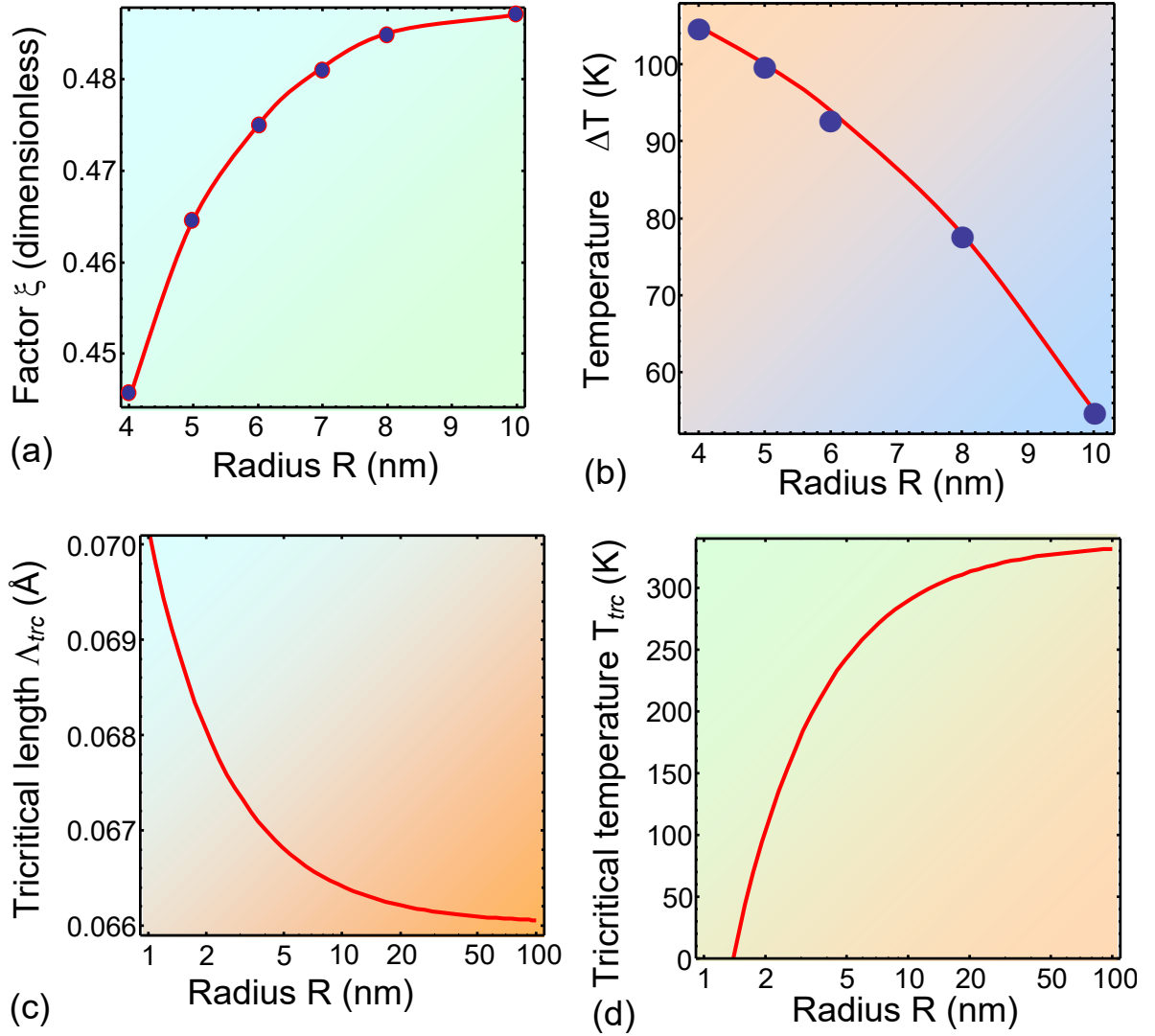

**FIG. 4.** Radius dependences of the dimensionless geometrical factor $\xi(R)$ **(a)** and temperature $\Delta T(R)$ **(b)**, tricritical screening length $\Lambda_{tcr}(R)$ **(c)** and tricritical temperature $T_{tcr}(R)$ **(d)** calculated at $g_{44}=10^{-11}$ m$^3$/F. The tricritical values have been calculated from Eqs.(8). Parameters $\xi(R)$ and $\Delta T(R)$ (symbols) and critical index $\delta = 3/2$ have been extracted from FEM results shown in **Appendix C**. Solid curves in plots **(a)** and **(b)** are guided by eye. SPS parameters are listed in **Table 1.**

Radius dependences of the screening length $\Lambda_{tcr}(R)$ and tricritical temperature $T_{tcr}(R)$ are shown in **Figs.4(c)-(d)**, respectively. The tricritical values have been calculated from exact analytical expressions (8). From **Fig. 4(c)** the value $\Lambda_{tcr}(R)$ very slightly decreases from 0.07 Å to 0.066 Å with $R$ increase from 1 nm to 100 nm, at that the saturation value 0.066 Å is reached for $R \geq 100$ nm as anticipated from Eq.(8b) for SPS parameters. From **Fig. 4(d)** the value $T_{tcr}(R)$ rapidly increases from 50 K to 320 K with R increase from 2 nm to 100 nm, then saturates and tends to $T_C$ for $R \geq 100$ nm as anticipated from Eq.(8a).

Note that the increase of $T_{tcr}(R)$ above room temperature occurring in SPS particles with $R$ increase of more than 50 nm is likely an artifact of the oversimplified LGD free energy expansion on 2-4-6 polarization powers given by Eq.(1b). Actually Eq.(1b) does not contain 8-th and 10-th powers of polarization, which are relevant for precise quantitative description of the Sn$_2$P$_2$(S,Se)$_6$ (see e.g. [65]). In fact, for a bulk sample under the normal pressure, there is the second-order phase transition, and the tricritical point is achieved by compressing (or replacing tin with lead) and lowering the temperature to 250 K [66]. Such a change in the transition order with the decreasing temperature is typical for the case of a three-well local potential (Blume-Emery-Griffiths model [67]), which can be taken into account in the Landau theory by adding the invariants of the 8-th and 10-th polarization powers [65]. Yevych et al also tried to describe the situation using the model of quantum anharmonic oscillators [65, 68]. When the shape of the three-well potential is changed under e.g. compression of the crystal, the transition becomes of the first-order and metastable states appear. If the model is supplemented with different ratios of the interactions between the first and the next neighbors [69], and this is equivalent to the ANNNI model [12], and there will be antiferroelectric ordering and metastable states. The inter-cells interactions can be more accurately considered by taking into account the gradient invariants of higher order in Eq.(1c).

## V. COMPARISON WITH EXPERIMENT

As it was argued in the introduction, derived analytical expressions should be compared with experimental results for nanoparticles of uniaxial ferroelectrics. Our first choice is SrBi$_2$Ta$_2$O$_9$ particles, for which the dependence of ferroelectric transition temperature on the particles' average radius was measured in Ref. [14] by XRD and *in situ* Raman scattering. The experimental data [14] is described by empirical Ishikawa equation [70],

$$T_{Ish}(R) = T_C \left(1 - \frac{\Delta R}{R - r_{cr}}\right), \quad (11)$$

with the bulk Curie temperature $T_C = 605$ K, critical radius $r_{cr} = 1.05$ nm and $\Delta R = 1.90$ nm [compare diamonds and dashed red curve in **Fig. 5(a)**]. One can see from **Fig. 5(a)** that the analytical expression (5) for the radius dependence of the poly-domain particle (black solid curve) perfectly reproduced empirical Ishikawa equation (dashed red curve) at all temperatures. At the same time the transition temperature calculated from Eq.(4) in the assumption that the particles are single-domain (dotted blue curve) is significantly smaller than the best fitting and thus the assumption does not describe experimental results adequately.

Formally Eqs.(5a) looks very different from Ishikawa equation (11). However both equations can be rewritten in a similar form for particle radii close to the critical value. Actually, Eq.(11) can be presented in the identical form as $T_{Ish}(R) = T_C \frac{R - R_{Ish}}{R - r_{cr}}$, where $R_{Ish} = r_{cr} + \Delta R$. For the particle radii $R \to R_{Ish}$ the Ishikawa formulae can be approximated as $T_{Ish}(R) \sim T_C \frac{R - R_{Ish}}{R}$. At $R > R_{cr}$ Eq. (5a) can be presented in the form, $T_{PE-PDFE}(R) = T_C \frac{(R - R_1)(R - R_2)}{R^2}$, where $R_1 = \frac{\Delta R_1}{2} - \sqrt{\frac{\Delta R_1^2}{4} - \Delta R_1 \Delta R_2}$ and $R_2 = \frac{\Delta R_1}{2} + \sqrt{\frac{\Delta R_1^2}{4} - \Delta R_1 \Delta R_2}$ are positive roots of the quadratic equation $R^2 - \Delta R_1 R + \Delta R_1 \Delta R_2 = 0$ with $\Delta R_1(\Lambda) = \frac{g_{44}}{\alpha_T T_C \xi} \left( \frac{1}{\sqrt{\varepsilon_0(\varepsilon_b + 2\varepsilon_e)g_{44}}} + \frac{1}{\xi R_{cr}} \right)$ and $\Delta R_2 = \frac{1}{\xi} \left( \frac{1}{\sqrt{\varepsilon_0(\varepsilon_b + 2\varepsilon_e)g_{44}}} + \frac{1}{\xi R_{cr}} \right)^{-1}$. When the particle radius is close to the smaller root of the quadratic equation, $R \to R_1$, the transition temperature can be approximated as $T_{PE-PDFE}(R) \sim T_C \frac{(R - R_1)}{R}$, and the latter expression is similar to Ishikawa expression, $T_{Ish}(R) \sim T_C \frac{R - R_{Ish}}{R}$, at $R \to R_{Ish}$. The similarity explains the proximity of solid and dashed curves in **Fig.5(a)**, and, more important, gives theoretical grounds to the empirical Ishikawa expression.

Since the previous section analysis phase diagrams and domain structure of the $Sn_2P_2S_6$ nanoparticles, we decided to compare the theoretical dependences the experimental ones. To the best of our knowledge, corresponding experiments are still absent, only MC simulations within the framework of ANNNI model was done in Ref. [12] for SPS nanocrystals. Nanocrystals have different sizes from 9 to 67 cell units (c.u.) and where imposed to the periodic boundary conditions (see Fig.6 in Ref.[12]). To recalculate the cell unit into a physical size for SPS, we used the pseudo-orthorhombic setting (see e.g. [71]) with a lattice parameters $a$ = 0.9318 nm, $b$ = 0.7463 nm and $c$ = 0.6518 nm at 358 K. Since the cubic cells was used in Ref.[12], we regarded that a cell unit parameter is approximately equal to $\sqrt[3]{abc}$ =0.7323 nm. Diamonds in **Fig. 5(b)** are MC ANNNI modeling simulation results from Ref. [12]. Dashed curve represents empirical Ishikawa equation $T_c = 337(1 - 1.48/(R - 0.4))$. Dotted and solid curves in **Fig. 5(b)** is the fitting with Eq.(4) (dotted blue curves) and Eq.(5) (solid black curves) for the material parameters of SPS, $T_c = 337$ K, $\alpha_T = 1.6 \times 10^6$ C$^{-2}$·mJ/K, $g_{44}=2.3\times10^{-12}$ m$^3$/F, $\varepsilon_b = 7$, geometrical factor $\xi \approx 0.50$ and surface screening length $\Lambda$=0.3 Å. As one can see from the **Fig. 5(b)** analytical expression (5a) for the radius dependence of the poly-domain particle (black solid curve) perfectly reproduced empirical Ishikawa equation (dashed red curve) at all temperatures. At the same time the transition temperature calculated from Eq.(4) in the assumption that the particles are single domain (dotted blue curve) is significantly lower than the best fitting and thus the single-domain assumption does not describe experimental results at all.

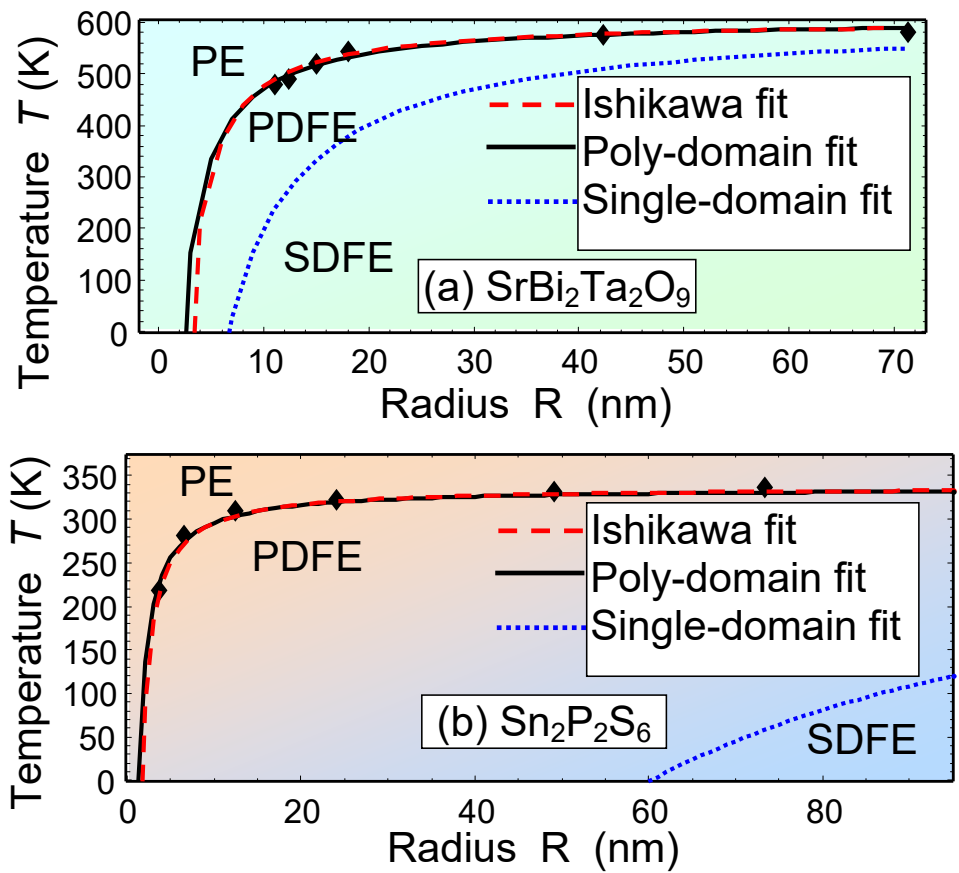

FIG. 5. (a) Dependence of PE-FE transition temperature on the radius of SrBi$_2$Ta$_2$O$_9$ particles. Diamonds are experimental data from Ref. [14]. Dotted blue curve is the fitting by Eq.(4) and solid black curve is the fitting by Eq.(5) for parameters $T_c = 608$ K, $\alpha_T = 4.06\times10^5$C$^{-2}$·mJ/K, $g_{44}=2.5\times10^{-12}$ m$^3$/F, $\varepsilon_b = 10$, $\xi \approx 0.5$ and $\Lambda$=0.15Å. Dashed curve is empirical Ishikawa equation $T_c = 605(1 - 1.90/(R - 1.05))$ with radius $R$ in nanometers. **(b)** Dependence of PE-FE transition temperature on the radius of Sn$_2$P$_2$S$_6$ particles. Diamonds are MC ANNNI modeling simulation results from Ref. [12]. Dotted blue curve is the fitting by Eq.(4) and solid black curve is the fitting by Eq.(5) for parameters $T_c = 337$ K, $\alpha_T = 1.6 \times 10^6$ C$^{-2}$·mJ/K, $g_{44} = 2.3 \times 10^{-12}$ m$^3$/F, $\varepsilon_b = 7$, $\xi \approx 0.50$ and $\Lambda = 0.3$ Å. Dashed curve is empirical Ishikawa equation $T_c = 337(1 - 1.48/(R - 0.4))$ with radius $R$ in nanometers.

Temperature dependence of the spontaneous polarization in the nanoparticles can be estimated from the expression

$$P_S(R, \Lambda, T) = \sqrt{\frac{1}{2\gamma}\left(\beta - \sqrt{\beta^2 + 4\gamma\alpha_T(T_{cr}(R,\Lambda) - T)}\right)} \approx \sqrt{-\frac{\alpha_T}{\beta}(T_{cr}(R,\Lambda) - T)} \quad (12)$$

Where $T_{cr}(R,\Lambda)$ is given by Eq.(4) for a single-domain case or estimated from Eqs.(5) for a poly-domain case in the sense of maximal value (because the average value is zero in a poly-domain particle).

Temperature dependence of the spontaneous polarization calculated by MC ANNNI model simulations in SPS particles with different radiuses $R$ is shown in **Fig. 6** by diamonds. Solid curves is the fitting with Eq.(12) for the same parameters as in **Fig.5(b)**. From **Fig.6** we conclude that Eq.(12) can describe semi-quantitatively the polarization temperature behavior in the vicinity of FE-PE phase transition, but not the polarization saturation far from the transition. Most likely the discrepancy is related with the principal differences between the continuous Landau free energy expansion on 2-4-6 polarization powers that is applicable to the ferroelectric phase transitions of displacement type and ANNNI model that rather describes the order-disorder phase transitions (see also remark at p.12 and Refs.[65-68]).

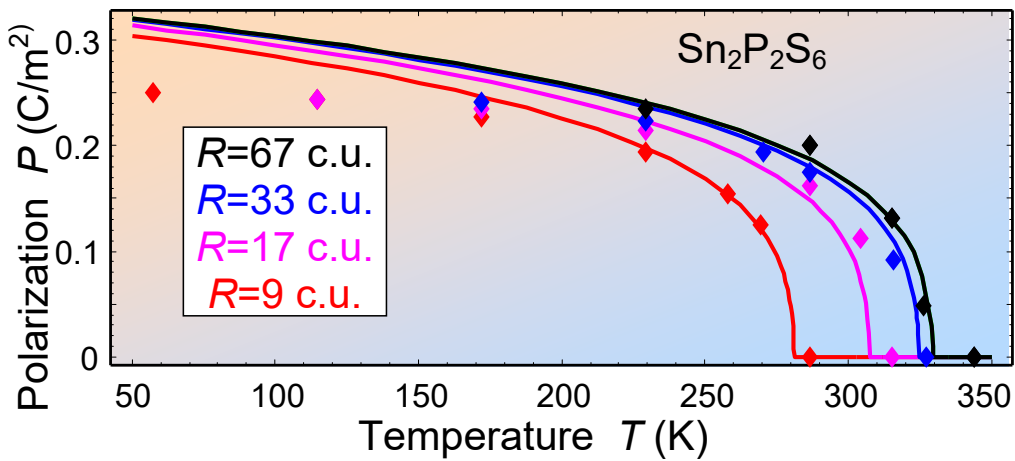

**FIG. 6.** Temperature dependence of the spontaneous ferroelectric polarization in the Sn$_2$P$_2$S$_6$ particles with different radiuses $R$ (in cell units described in the text). Diamonds are MC ANNNI model simulations from Ref. [12]. Solid curves is the fitting with Eq.(12) for the same parameters as in **Fig.5(b)**. A cell unit (**c.u.**) parameter is approximately equal to 0.7323 nm.

# VI. DISCUSSION

In the framework of LGD approach we evolved analytical description of domain structure morphology and phase diagrams of ferroelectric nanoparticles covered by an ultra-thin layer of screening charge characterized by finite screening length $\Lambda$. The phase diagrams, calculated by FEM in coordinates "temperature $T$ – surface screening length $\Lambda$" for spherical $Sn_2P_2S_6$ nanoparticles with radius $R$, demonstrate the appearance of poly-domain region in the tricritical point and its broadening with $\Lambda$ increase above 0.1 Å and $R$ increase above 4 nm. Typically the poly-domain ferroelectric region (PDFE) of triangular-like shape separates single-domain ferroelectric (SDFE) and paraelectric (PE) phases.

Metastable labyrinthine domains (LD) region was revealed in 10-nm $Sn_2P_2S_6$ nanoparticles, but absolutely stable labyrinthine domains are likely absent in contrast to the case of $CuInP_2S_6$ nanoparticles considered previously [45]. LD are expected to be a long-living stable (or at least metastable) configuration in $Sn_2P_2S_6$ nanoparticles undergoing the second order phase transition at Curie temperature. Earlier the LD structure was predicted theoretically in thin films of incommensurate ferroelectrics [72] and bi-layered ferroelectrics [73], being similar to those observed in ultrathin magnetic films [74]. Notably, fractal domain structures are sometimes observed in multiferroic thin films [75] and near the surface of relaxors close to relaxor-ferroelectric boundary [76], but the stable LD with a single characteristic length scale were observed by Vectorial Piezoresponse Force Microscopy (V-PFM) [77] in ergodic relaxors only [78]. Metastable labyrinthine domains can coexist with classical ferroelectric domains [79, 80].

That say, our theoretical prediction of either stable or metastable LD appearance requires experimental verification. V-PFM can be an appropriate method for the 3D visualization of the labyrinthine domain structure with nanoscale resolution (see e.g. [77, 78, 81, 82, 83, 84, 85] and refs therein), because stable surface-induced labyrinthine domain structures were revealed by PFM in ferroelectrics relaxors [78]. Noteworthy different domain morphologies inside the small nanoparticles ($R < 10$ nm) can be probed by abbreviation corrected high resolution scanning transmission electron microscopy (STEM) [3 - 5].

We leave for further studies the question how the ranges of LDs stability and metastability at phase diagram can be derived analytically. However, the necessary conditions of LDs appearance was derived analytically with the numerical factor estimated from FEM results. Obtained analytical results lead to the conclusion that the nature of morphological transition [45] from domain stripes to LDs in ferroelectric nanoparticles underlies in the interplay between the 2D instability of domain splitting induced by incomplete surface screening facilitated by small enough gradient energy and high enough particle size to prevent the effect of geometric catastrophe. Quantitative criteria of LDs appearance has the form $\frac{L_d}{\Lambda} \leq \mu$, where $L_d = \sqrt{\varepsilon_0 g_{44}}$ is the depolarization length and $\mu = 2\pi\sqrt{\varepsilon_b + 2\varepsilon_e}$ stands for effective "geometrical factor". The physical sense of the condition is that the ratio of depolarization to screening lengths should be smaller than.

Also we derived analytical expressions for all other boundaries at the phase diagrams and checked their accuracy by comparison with numerical results. Notably that the simple analytical expression (4) for the PE-SDFE transition temperature, that is exact for the natural boundary conditions at the particle surface, depends on the ratio $(R/\Lambda)$ only. The relatively simple analytical expressions (5) for the PE-PDFE transition temperature appear to have very high accuracy and contain only one fitting parameter $\xi$ that's value $\xi \approx 0.5$ is conditioned by the spherical geometry of the particle. Since the value $\xi \approx 0.5$ is not material specific, Eq.(5) can be used for description of PE-PDFE transition in other nanoparticles of uniaxial ferroelectrics (e.g. for $SrBi_2Ta_2O_9$, $CuInP_2S_6$ or $LiNbO_3$). Analytical expression (9) quantitatively describes the boundary between SDFE and PDFE phases with two extra fitting parameters, the temperature shift $\Delta T$ and critical index $\delta=3/2$, which depend on the material parameters, and therefore this expression is not universal. Analytical expressions (8) for the dependence of the critical point temperature $T_{tcr}(R)$ and screening length $\Lambda_{tcr}(R)$ on the particle radius are almost exact and universal showing that the dependences $T_{tcr}(R)$ and $\Lambda_{tcr}^{-1}(R)$ scales as $1/R$ for ferroelectric nanoparticles.

The derived analytical expressions were compared with the dependence of FE-PE transition temperature on the radius of $SrBi_2Ta_2O_9$ nanoparticles experimentally measured in Ref.[14], and with analogous dependence for $Sn_2P_2S_6$ nanocrystals simulated by MC method within the framework of ANNNI model in Ref. [12]. Comparison demonstrates that the analytical expression (5) for the radius dependence of the FE-PE transition temperature in a poly-domain particle quantitatively agrees with experimental [14] and simulated [12] results, as well as it perfectly reproduces empirical Ishikawa equation at all temperatures. More important, the analytical expression gives theoretical grounds to the empirical Ishikawa expression. At the same time the PE-FE transition temperature calculated from Eq.(4) in the assumption that the particles are single domain does not describe experimental results adequately.

Temperature dependence of the $Sn_2P_2S_6$ nanocrystalls spontaneous polarization was calculated by Drobnich et al using MC method and ANNNI model [12]. Derived analytical expression (12) can describe semi-quantitatively the polarization temperature behavior in the vicinity of FE-PE phase transition, but not its saturation far from the transition. Most likely the discrepancy is related with the principal differences between the continuous Landau free energy expansion on polarization powers that is applicable to the ferroelectric phase transitions of displacement type and ANNNI model that rather describes the order-disorder phase transitions.

To resume, we can conclude that phase diagrams including the wide regions of striped and labyrinthine domain morphologies, which are qualitatively similar to the ones calculated in this work, can be realized in other uniaxial ferroelectric nanoparticles, such as $SrBi_2Ta_2O_9$ and $LiNbO_3$, with the sizes near the PE-FE transition. Notably, the domain morphologies and phase diagrams of nanoparticles are rather different for the uniaxial ferroelectrics with the first and second order PE-FE transitions. Much more complex situation (corresponding to the balance of labyrinthine domains in the bulk and vortices at the surface) are expected in multiaxial ferroelectric nanoparticles with polarization rotation allowed. Actually, vortices and vertices composed by the closure of four domain walls have been observed experimentally by STEM [3 - 5, 86] and PFM [87, 88, 89], and described theoretically [90] in a bulk and nanosized multiaxial ferroelectrics.

**Acknowledgements.** This project has received funding from the European Union's Horizon 2020 research and innovation programme under the Marie Skłodowska-Curie grant agreement No 778070. A.N.M. work was partially supported by the National Academy of Sciences of Ukraine (project No. 0118U003375 and No. 0117U002612) and by the Program of Fundamental Research of the Department of Physics and Astronomy of the National Academy of Sciences of Ukraine (project No. 0117U000240). P.M. was supported by the Center for Nanophase Materials Sciences, sponsored by the Division of User Facilities, Basic Energy Sciences, US Department of Energy.

**Authors' contribution.** A.N.M. generated research idea, stated the problem, derived analytical expressions jointly with E.A.E., interpreted results, compared with experiment and MC simulations and wrote the manuscript. E.A.E. wrote the codes and performed numerical calculations. Y.M.F. tested the codes and assisted E.A.E. with simulations. P.M. and Yu.M.V. densely worked on the results discussion and manuscript improvement.

# SUPPLEMENTARY MATERIALS

## Appendix A. Derivation of PE-SDFE transition temperature

Let us consider the spherical ferroelectric particle with polarization **P** pointed along one of the principal crystallographic axis, denoted below as $z$. Here we also introduce an isotropic background permittivity $\varepsilon_b$ of ferroelectric. The media outside is a dielectric with permittivity $\varepsilon_e$. Electrical displacement is $\mathbf{D}_i = \varepsilon_0 \varepsilon_b \mathbf{E}_i + \mathbf{P}$ and $\mathbf{D}_e = \varepsilon_0 \varepsilon_e \mathbf{E}_e$, where the subscript "$i$" means the physical quantity inside the particle, "$e$" – outside the particle; $\varepsilon_0$ is a universal dielectric constant. We introduce electric field $\mathbf{E} = -\nabla\varphi$ via electrostatic potential $\varphi$, which should satisfy Poisson and Laplace equations inside and outside the particle, respectively

$$\varepsilon_0 \varepsilon_b \left( \frac{\partial^2}{\partial x^2} + \frac{\partial^2}{\partial y^2} + \frac{\partial^2}{\partial z^2} \right) \varphi_i = -\frac{\partial P}{\partial z} \quad \text{inside particle} \qquad (A.1a)$$

$$\varepsilon_0 \varepsilon_e \left( \frac{\partial^2}{\partial x^2} + \frac{\partial^2}{\partial y^2} + \frac{\partial^2}{\partial z^2} \right) \varphi_e = 0 \quad \text{outside particle} \qquad (A.1b)$$

supplemented by the interface conditions of potential continuity at the particle surface S both for potential and normal components of electrical displacement:

$$\left( \varphi_e - \varphi_i \right)\big|_S = 0; \qquad \left( (\mathbf{D}_e - \mathbf{D}_i)\mathbf{n} + \varepsilon_0 \frac{\varphi_i}{\Lambda} \right)\bigg|_S = 0 \qquad (A.2)$$

Here **n** is the outer normal to the particle surface. In Eq.(A.2) we take into consideration the surface screening charge density, proportional to the surface potential. $\Lambda$ is the surface screening length. Below we suppose that polarization **P** is independent on the coordinates; there is homogeneous external electric field $E_0$ far from the particle. This problem is reminiscent of text-book ones (see e.g. [V. V. Batygin, and I. N. Toptygin. "Problems in Electrodynamics (Academic, New York, 1978)."]) but without surface screening charge, which makes the problem solution non-trivial.

For a spherical particle the general solutions of Eqs.(A.1) could be expanded into the series on Legendre polynomials. For the considered problem few terms are sufficient, namely

$$\varphi_e = -E_0 r \cos\theta + E_e \frac{R^{s+1}}{r^s} \cos\theta, \qquad \varphi_i = -E_i r \cos\theta. \qquad (A.3)$$

Here $\theta$ is the polar angle for spherical coordinate system, $r$ is the corresponding radial coordinate. $s = 1$ and $s = 2$ are for and cylindrical polar and spherical coordinate systems respectively. $E_e$ and $E_i$ are the constants to be determined from the boundary conditions (A.2).

Application of Eqs.(A.2) to (A.3) leads to condition

$$-E_0 R\cos\theta + E_e \frac{R^{s+1}}{R^s}\cos\theta = -E_i R\cos\theta \quad \Rightarrow \quad E_e + E_i = E_0 \qquad (A.4)$$

Radial components of field could be obtained from (A.3) as follows

$$(\mathbf{E}_e)_r = E_0\cos\theta + sE_e\frac{R^{s+1}}{r^{s+1}}\cos\theta, \qquad (\mathbf{E}_i)_r = E_i\cos\theta. \qquad (A.5)$$

Corresponding displacement is

$$(\mathbf{D}_e)_r = \varepsilon_0\varepsilon_e\left(E_0\cos\theta + sE_e\frac{R^{s+1}}{r^{s+1}}\cos\theta\right), \qquad (\mathbf{D}_i)_r = \varepsilon_0\varepsilon_b E_i\cos\theta + P\cos\theta \qquad (A.6)$$

Application of Eq.(A.2) to Eq.(A.3) and (A.6) leads to conditions

$$\varepsilon_0\varepsilon_e(E_0\cos\theta + sE_e\cos\theta) - \varepsilon_0\varepsilon_b E_i\cos\theta - P\cos\theta - \varepsilon_0\frac{E_i R\cos\theta}{\Lambda} = 0. \quad \Rightarrow$$

$$\varepsilon_e s E_e - \left(\varepsilon_b + \frac{R}{\Lambda}\right) E_i = \frac{P}{\varepsilon_0} - \varepsilon_e E_0 \qquad (A.7)$$

The solution of the linear system (A.4) and (A.7) have the form:

$$E_i = -\frac{P}{\varepsilon_0}\frac{1}{\varepsilon_b + \varepsilon_e s + (R/\Lambda)} + \frac{\varepsilon_e(1+s)}{\varepsilon_b + \varepsilon_e s + (R/\Lambda)}E_0, \quad E_e = \frac{P}{\varepsilon_0}\frac{1}{\varepsilon_b + \varepsilon_e s + (R/\Lambda)} + \frac{\varepsilon_b - \varepsilon_e + (R/\Lambda)}{\varepsilon_b + \varepsilon_e s + (R/\Lambda)}E_0$$

(A.8)

with s=2 for a sphere and s=1 for a cylinder. Below we use the expression (A.8) for the formulation of the phenomenological equations of state. Including internal electric field (A.8) into the LGD equation one obtain for the sphere

$$\left(\alpha_T(T-T_c) + \frac{1}{\varepsilon_0(\varepsilon_b + 2\varepsilon_e + (R/\Lambda))}\right)P + \beta P^3 = \frac{3\varepsilon_e E_0}{(\varepsilon_b + 2\varepsilon_e + (R/\Lambda))} \qquad (A.9a)$$

### Appendix B. Derivation of PE-PDFE transition temperature

The linearized system of equations for polarization and electric potential inside the ferroelectric nanoparticle and outside it has the following form

$$\alpha P_3 - g_{11}\frac{\partial^2 P_3}{\partial z^2} - g_{44}\left(\frac{\partial^2 P_3}{\partial x^2} + \frac{\partial^2 P_3}{\partial y^2}\right) = -\frac{\partial\varphi}{\partial z}, \qquad (B.1a)$$

$$\left(\frac{\partial^2}{\partial x^2}+\frac{\partial^2}{\partial y^2}+\frac{\partial^2}{\partial z^2}\right)\varphi^{(in)}=\frac{1}{\varepsilon_0\varepsilon_b}\frac{\partial P_3}{\partial z}, \quad \left(\frac{\partial^2}{\partial x^2}+\frac{\partial^2}{\partial y^2}+\frac{\partial^2}{\partial z^2}\right)\varphi^{(out)}=0, \qquad (B.1b)$$

with appropriate boundary conditions at the particle surface S:

$$\left(\frac{\partial P_3}{\partial r}\right)\bigg|_{r=R}=0, \quad \left(-\varepsilon_0\varepsilon_b\frac{\partial\varphi^{(in)}}{\partial r}+P_3+\varepsilon_0\varepsilon_e\frac{\partial\varphi^{(out)}}{\partial r}-\varepsilon_0\frac{\varphi^{(out)}}{\Lambda}\right)\bigg|_{r=R}=0. \quad \left(\varphi^{(out)}-\varphi^{(in)}\right)\bigg|_{r=R}=0 \quad (B.2)$$

Let us consider harmonic like fluctuations

$$P_3=P_k(z)\exp(i\vec{k}\vec{r}), \quad \varphi^{(in)}=\varphi_k^{(in)}(z)\exp(i\vec{k}\vec{r}), \quad \varphi^{(out)}=\varphi_k^{(out)}(z)\exp(i\vec{k}\vec{r}). \qquad (B.3)$$

Equations for amplitudes

$$\left(\alpha+g_{44}k^2\right)P_k-g_{11}\frac{\partial^2 P_k}{\partial z^2}=-\frac{\partial\varphi_k}{\partial z}, \qquad (B.3a)$$

$$\frac{\partial^2\varphi_k^{(in)}}{\partial z^2}-k^2\varphi_k^{(in)}=\frac{1}{\varepsilon_0\varepsilon_b}\frac{\partial P_k}{\partial z}, \qquad \frac{\partial^2\varphi_k^{(out)}}{\partial z^2}-k^2\varphi_k^{(out)}=0. \qquad (B.3b)$$

Where $k^2=k_x^2+k_y^2$. Differentiation the Eqs. (B.3) gives

$$\left(\frac{\partial^2}{\partial z^2}-k^2\right)\left[\left(\alpha+g_{44}k^2\right)P_k-g_{11}\frac{\partial^2 P_k}{\partial z^2}\right]=\left(\frac{\partial^2}{\partial z^2}-k^2\right)\left[-\frac{\partial\varphi_k}{\partial z}\right], \qquad (B.4a)$$

$$\frac{\partial}{\partial z}\left(\frac{\partial^2}{\partial z^2}-k^2\right)\varphi_k^{(in)}=\frac{1}{\varepsilon_0\varepsilon_b}\frac{\partial^2 P_k}{\partial z^2}. \qquad (B.4b)$$

Hence, one could exclude the potential amplitude from Eq. (B.4a) and get the single equation for polarization amplitude in the form:

$$\left(\frac{\partial^2}{\partial z^2}-k^2\right)\left[\left(\alpha+g_{44}k^2\right)P_k-g_{11}\frac{\partial^2 P_k}{\partial z^2}\right]=-\frac{1}{\varepsilon_0\varepsilon_b}\frac{\partial^2 P_k}{\partial z^2} \qquad (B.5)$$

Let us look for the solution of (B.5) in the form $P_3 \sim \exp(qz)$, where inverse characteristic length $w$ satisfies the following equation:

$$q^4-\left(\frac{\alpha+g_{44}k^2}{g_{11}}+k^2+\frac{1}{\varepsilon_0\varepsilon_b g_{11}}\right)q^2+\frac{\alpha+g_{44}k^2}{g_{11}}k^2=0 \qquad (B.6a)$$

Its solutions could be written as

$$q_{1,2}^2 = \frac{1}{2}\left( \frac{\alpha + g_{44}k^2}{g_{11}} + k^2 + \frac{1}{\varepsilon_0\varepsilon_b g_{11}} \pm \sqrt{\left(\frac{\alpha + g_{44}k^2}{g_{11}} + k^2 + \frac{1}{\varepsilon_0\varepsilon_b g_{11}}\right)^2 - 4\frac{\alpha + g_{44}k^2}{g_{11}}k^2} \right) \quad \text{(B.6b)}$$

It should be noted, that in the most cases $\varepsilon_0\varepsilon_b g_{11} \ll \{1/k^2, g_{11}/|\alpha|, g_{44}/|\alpha|\}$, hence the following approximations are valid

$$q_1 \approx \sqrt{\frac{(\alpha + g_{44}k^2)k^2}{\alpha + g_{44}k^2 + g_{11}k^2 + \frac{1}{\varepsilon_0\varepsilon_b}}}, \quad q_2 \approx \frac{1}{\sqrt{\varepsilon_0\varepsilon_b g_{11}}} \quad \text{(B.6c)}$$

Now we could write the general solution of Eq.(B.5) in the form:

$$P_3 = s_1 \sinh(q_1 z) + s_2 \sinh(q_2 z) + c_1 \cosh(q_1 z) + c_2 \cosh(q_2 z) \quad \text{(B.7a)}$$

The four constants $s_i$ and $c_i$ should be found from boundary conditions (B.2). Formal solution is zero, since we have the system of homogeneous linear equations for $s_i$ and $c_i$, but we are interested in the stability analysis, hence we should look for zero point of the corresponding determinant of the linear equations system for $s_i$ and $c_i$. Since counter domain walls are charged and hence have much higher energy in comparison with parallel ones, the antisymmetric part of the solution corresponding to nonzero $s_i$ is always unstable from energetic considerations.

Unfortunately exact solution for the constants $c_i$ in the functions $P_3 = c_1 \cosh(q_1 z) + c_2 \cosh(q_2 z)$ and constants $f_i$, $f$ in electric potentials $\varphi_k^{(in)} = f_1 \sinh(q_1 z) + f_2 \sinh(q_2 z)$ and $\varphi_k^{(out)} = f \exp(-|k|(z-h))$ are impossible to find in a finite form. Assuming that in the vicinity of the particle poles $z = \pm R$, the curvature of the spherical surface can be neglected, we obtained the system of four linear equations

$$c_1 q_1 \sinh(q_1 h) + c_2 q_2 \sinh(q_2 h) = 0, \quad f_1 \sinh(q_1 h) + f_2 \sinh(q_2 h) - f = 0, \quad c_i = \varepsilon_0\varepsilon_b \frac{q_i^2 - k^2}{q_i} f_i \quad \text{(B.8a)}$$

$$-\varepsilon_0\varepsilon_b(q_1 f_1 \cosh(q_1 h) + q_2 f_2 \cosh(q_2 h)) + (c_1 \cosh(q_1 h) + c_2 \cosh(q_2 h)) - |k|\varepsilon_0\varepsilon_e f - \varepsilon_0 \frac{f}{\Lambda} = 0. \quad \text{(B.8b)}$$

After cumbersome calculations of the conditions of its zero determinant can be simplified under the validity of strong inequalities $|k|\varepsilon_e\Lambda \ll 1$ valid for most cases, therefore, recalling the condition $q_2 \gg |k|$,

$$\alpha + g_{44}k^2 + \frac{1}{\varepsilon_0(\varepsilon_b + (R/\Lambda) + \varepsilon_b(\xi k R)^2)} \approx 0 \quad \text{(B.9a)}$$

This expression for the critical point should be further minimized with respect to wave vector k.

$$\left(g_{44} - \frac{(\xi R)^2}{\varepsilon_0 \varepsilon_b \left(1 + \varepsilon_b^{-1}(R/\Lambda) + (\xi R \cdot k)^2\right)^2}\right) 2k = 0 \qquad (B.9a)$$

Since zero root $k=0$ of Eq.(B.9) corresponds to single domain sate, we neglected it and obtained the following value of domain structure wave vector at the transition point

$$k_{\min}(R,\Lambda) = \sqrt{\frac{1}{\sqrt{\varepsilon_0(\varepsilon_b + 2\varepsilon_e)g_{44}}\,\xi R} - \left(1 + \frac{R}{(\varepsilon_b + 2\varepsilon_e)\Lambda}\right)\left(\frac{1}{\xi R}\right)^2} \qquad (B.10)$$

It is valid under the condition $\dfrac{\xi}{\sqrt{\varepsilon_0(\varepsilon_b + 2\varepsilon_e)g_{44}}} \geq \dfrac{1}{R} + \dfrac{1}{(\varepsilon_b + 2\varepsilon_e)\Lambda}$. With respect to Eq. (B.10), Eq.(B.8) could be further simplified to $\alpha + \dfrac{\sqrt{\varepsilon_0 \varepsilon_b g_{44}}}{\varepsilon_0 \varepsilon_b} \dfrac{2}{\xi R} - g_{44}\left(1 + \dfrac{R}{\varepsilon_b \Lambda}\right)\left(\dfrac{1}{\xi R}\right)^2 \approx 0$.

Hence approximate analytical expression for the transition temperature of the spherical nanoparticle from the PDFE to PE phase is:

$$T_{PE-PDFE}(R,\Lambda) \approx T_C - \frac{1}{\alpha_T}\left(g_{44} k_{\min}^2(R,\Lambda) + \frac{\varepsilon_0^{-1}}{(R/\Lambda) + (\varepsilon_b + 2\varepsilon_e)\left(1 + k_{\min}^2(R,\Lambda)(\xi R)^2\right)}\right). \qquad (B.11a)$$

Here the first term originated from the correlation effect and the second one is from depolarization field energy of the domain stripes. Parameter $\xi$ is a sort of geometrical factor that is close to unity. Its origin is related to the corresponding spherical eigen functions. Corresponding wave vector $k_{\min}$ and period $D_{\max}$ of the domain structure onset are

$$k_{\min}(R,\Lambda) = \sqrt{\frac{1}{\sqrt{\varepsilon_0(\varepsilon_b + 2\varepsilon_e)g_{44}}\,\xi R} - \left(1 + \frac{R}{(\varepsilon_b + 2\varepsilon_e)\Lambda}\right)\left(\frac{1}{\xi R}\right)^2}, \qquad D_{\max} = \frac{2\pi}{k_{\min}}. \qquad (B.11b)$$

Substitution of Eq.(B.11b) into Eq.(B.11a) and elementary transformations leads to the evident expression for the PE-PDFE transition temperature radius dependence,

$$T_{PE-PDFE}(R,\Lambda) \approx \begin{cases} T_C - \dfrac{g_{44}}{\alpha_T \xi R}\left(\dfrac{1}{\sqrt{\varepsilon_0(\varepsilon_b + 2\varepsilon_e)g_{44}}} + \dfrac{1}{\xi R_{cr}(\Lambda)} - \dfrac{1}{\xi R}\right), & R > R_{cr} \\ 0, & R < R_{cr} \end{cases} \qquad (B.11c)$$

Where $R_{cr}(\Lambda) = \left(\dfrac{\xi}{\sqrt{\varepsilon_0(\varepsilon_b + 2\varepsilon_e)g_{44}}} - \dfrac{1}{(\varepsilon_b + 2\varepsilon_e)\Lambda}\right)^{-1}$ and $k_{\min}(R,\Lambda) = \dfrac{1}{\xi R}\sqrt{\dfrac{R}{R_{cr}(\Lambda)} - 1}$. Expressions (B.11) are valid under the condition

$$\dfrac{\xi}{\sqrt{\varepsilon_0(\varepsilon_b + 2\varepsilon_e)g_{44}}} \geq \dfrac{1}{R} + \dfrac{1}{(\varepsilon_b + 2\varepsilon_e)\Lambda} \qquad (B.12)$$

That means that the critical value of the gradient coefficient exists at fixed other parameters, $g_{44}^{cr}(R,\Lambda) = \dfrac{1}{\varepsilon_0(\varepsilon_b + 2\varepsilon_e)}\left(\dfrac{1}{\xi R} + \dfrac{1}{\xi(\varepsilon_b + 2\varepsilon_e)\Lambda}\right)^{-2}$, and domains appears at $g_{44} < g_{44}^{cr}(R,\Lambda)$. At fixed gradient coefficient $g_{44}$ the equality in Eq.(B.12) means that the relation between the particle radius R and screening length $\Lambda$ should be valid for the domain onset. Namely the fulfillment of the equality

$$\dfrac{1}{R} + \dfrac{1}{(\varepsilon_b + 2\varepsilon_e)\Lambda} = \dfrac{\xi}{\sqrt{\varepsilon_0(\varepsilon_b + 2\varepsilon_e)g_{44}}}. \qquad (B.13)$$

corresponding to the transition to a single domain state that occurs in a three-critical point on the phase diagram in coordinates e.g. *T* and $\Lambda$.

The stability of a PDFE state in comparison with a SDFE state depends on the balance between depolarization field energy (appearing from incomplete screening of the spontaneous polarization by surface charges with finite screening length $\Lambda$) and the domain walls energy (proportional to the gradient coefficient $g_{44}$). The domain splitting starts when it becomes energetically preferable. On the threshold of the domain formation the energy of SDFE state $G_{SD}$ includes the sum of electrostatic energy (1d) and Landau energy (1b) of a homogeneously polarized nanoparticle, $G_{SD} = G_{el} + G_{Landau} \to -\dfrac{\alpha_R^2}{4\beta}V$ (where $V = \dfrac{4\pi}{3}R^3$ and $\alpha_R = \alpha_T(T - T_{SDFE-PE}(R))$) becomes equal the energy of the particle with domain walls. The gradient energy (1c) is formally equal to the energy of the domain walls, $G_{grad} = G_S = \psi_S S_{DW}$, where $\psi_S$ is the surface energy of the domain wall and $S_{DW}$ is the area of domain walls. For infinitely thin domain walls $\psi_S$ is related with LGD coefficients as $\psi_S = \dfrac{2\sqrt{-2\alpha_R^3 g_{44}}}{3\beta}$ [91]. In accordance with Virial's theorem, the equality $|G_{SD}| = G_S$ leads to the equality $\dfrac{\alpha_R^2}{4\beta}V = \dfrac{2\sqrt{-2\alpha_R^3 g_{44}}}{3\beta}S_{DW}$ that's identical form is $T_{PDFE-SDFE} = T_{SDFE-PE}(R) - \dfrac{2g_{44}}{\alpha_T}\left(\dfrac{8S_{DW}}{3V}\right)^2$

**Appendix C. Phase diagrams of nanoparticles for different radius**

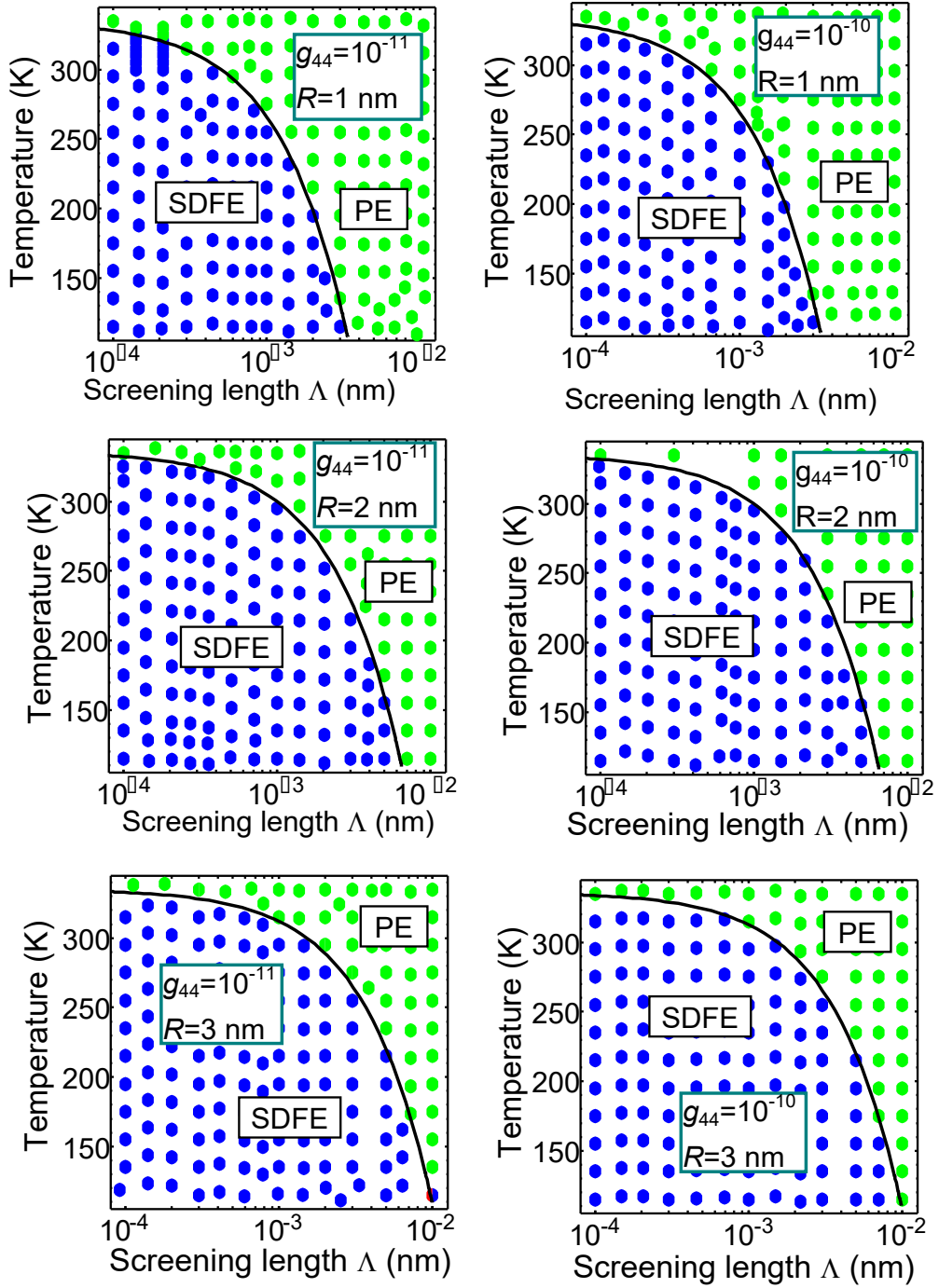

**FIG. S1.** Phase diagram of SPS nanoparticles in coordinates "temperature T – screening length Λ" calculated for different particle radius 1, 2 and 3 nm, gradient coefficient $g_{44}=10^{-11}$ m$^3$/F **(a)** and $g_{44}=10^{-10}$ m$^3$/F **(b)**. The ferroelectric single domain (SDFE) and paraelectric (PE) phases are stable. Solid curves corresponding to the phase SDFE-PE, boundaries are calculation from Eq.(4).

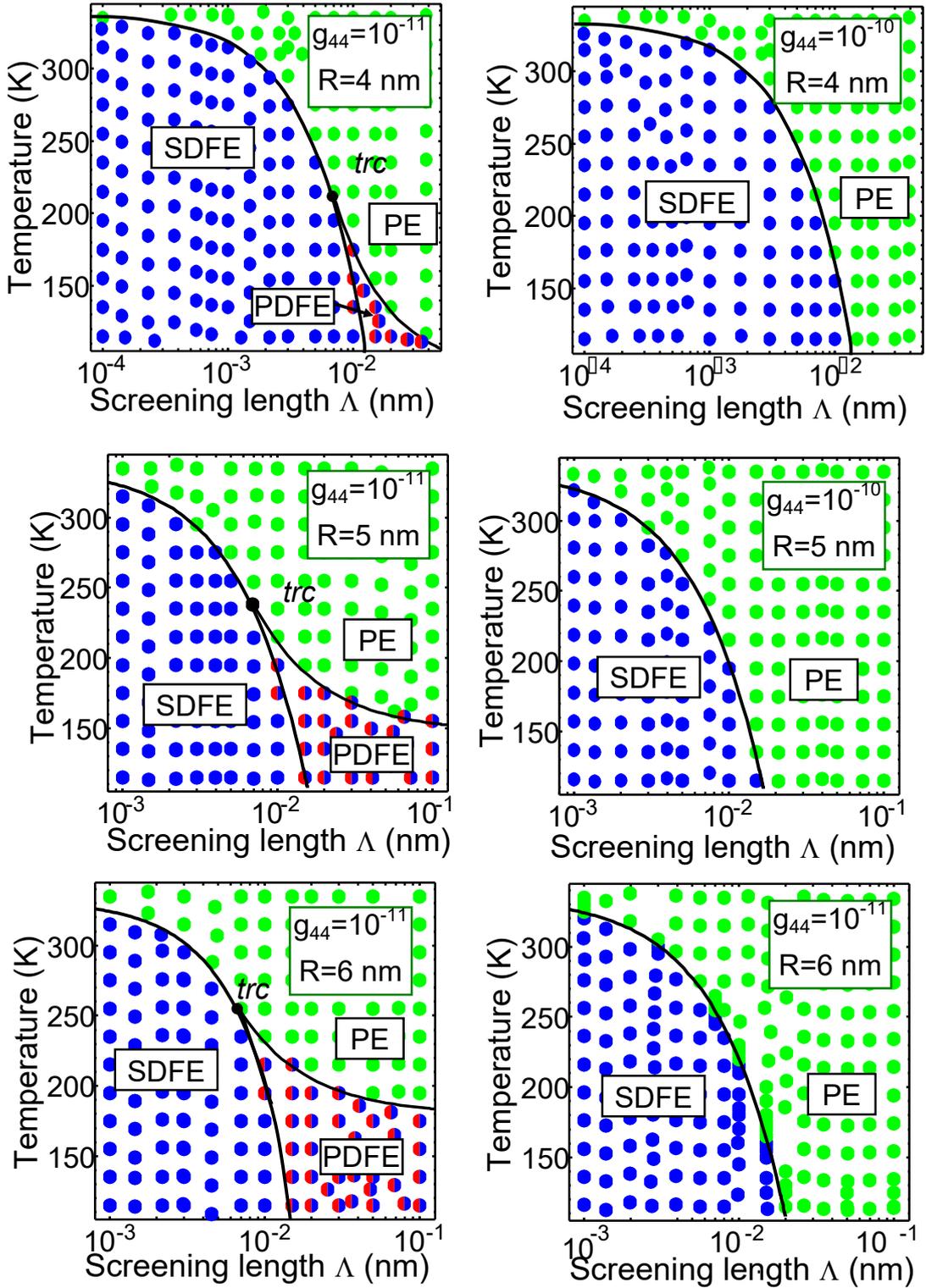

**FIG. S2.** Phase diagram of SPS nanoparticles in coordinates "temperature T – screening length Λ" calculated for different particle radius 4, 5 and 6 nm, gradient coefficient $g_{44}=10^{-11}$ m$^3$/F **(a)** and $g_{44}=10^{-10}$ m$^3$/F **(b)**. The ferroelectric single domain (SDFE), ferroelectric poly domain (PDFE) and paraelectric (PE) phases are stable. Solid curves corresponding to the SDFE-PE, PDFE-PE and SDFE-PDFE phase boundaries are calculated from Eqs.(4), (5) and (9), respectively.

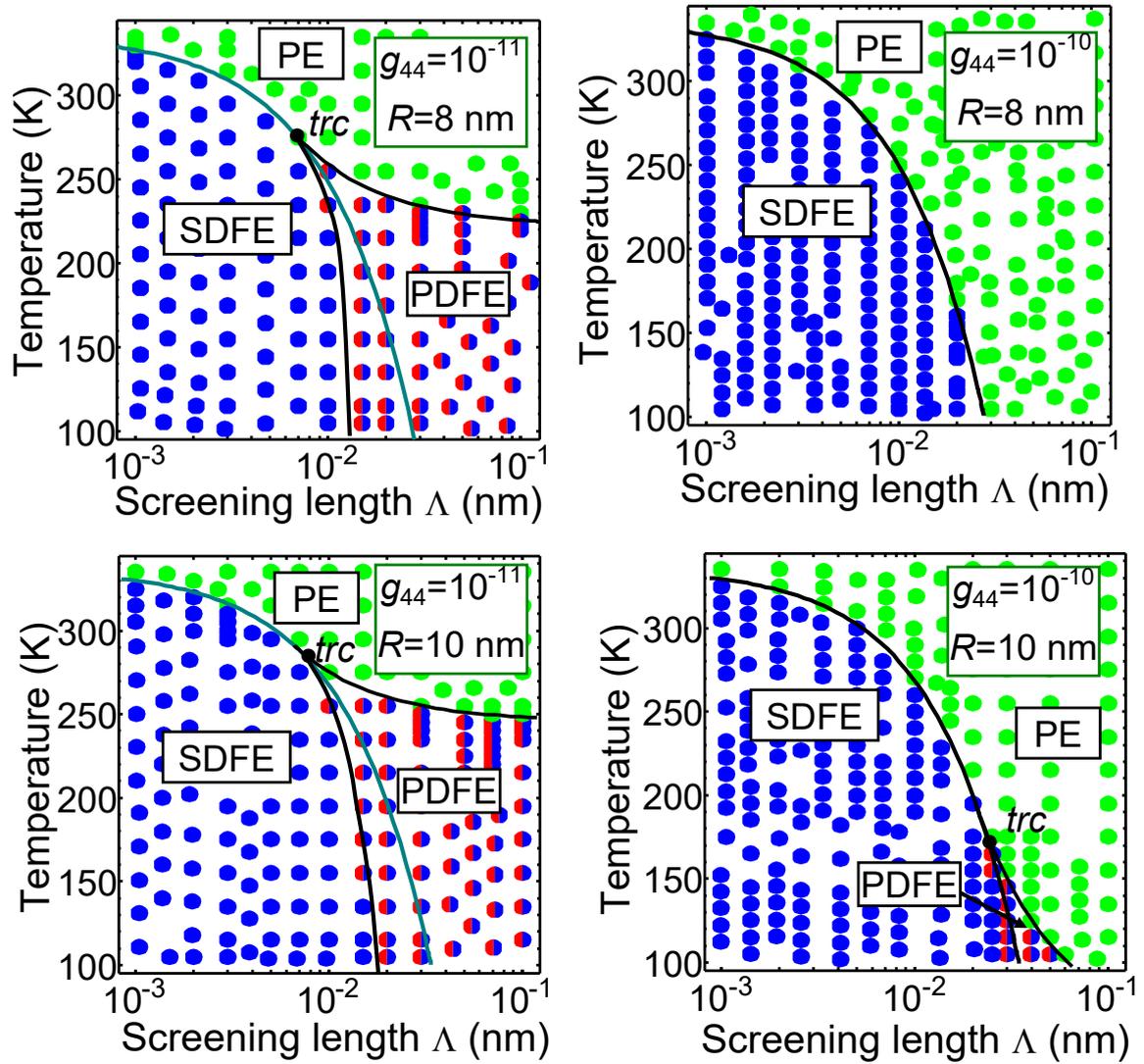

**FIG. S3.** Phase diagram of SPS nanoparticles in coordinates "temperature T – screening length Λ" calculated for different particle radius 8 and 10 nm, gradient coefficient $g_{44}=10^{-11}$ m$^3$/F **(a)** and $g_{44}=10^{-10}$ m$^3$/F **(b)**. The ferroelectric single domain (SDFE), ferroelectric poly domain (PDFE) and paraelectric (PE) phases are stable. Solid curves corresponding to the SDFE-PE, PDFE-PE and SDFE-PDFE phase boundaries are calculated from Eqs.(4), (5) and (9), respectively.

## REFERENCES

1 Topological Structures in Ferroic Materials. Springer Series in Mater. Sci. **228** 181-197 (2016). (DOI 10.1007/978-3-319-25301-5_8)

2 S. V. Kalinin, Y. Kim, D. Fong, and A.N. Morozovska, Surface-screening mechanisms in ferroelectric thin films and their effect on polarization dynamics and domain structures, Reports on Progress in Physics **81**, 036502 (2018).

3 Y. L. Tang, Y. L. Zhu, X. L. Ma, A. Y. Borisevich, A. N. Morozovska, E. A. Eliseev, W. Y. Wang Y. J. Wang, Y. B. Xu, Z. D. Zhang, S. J. Pennycook. Observation of a periodic array of flux-closure quadrants in strained ferroelectric PbTiO3 films. Science **348**, no. 6234, 547-551 (2015) 4 Sirui Zhang, Xiangwei Guo, Yunlong Tang, Desheng Ma, Yinlian Zhu, Yujia Wang, Shuang Li, Mengjiao Han, Dong Chen, Jinyuan Ma, Bo Wu, and Xiuliang Ma. "Polarization Rotation in Ultrathin Ferroelectrics Tailored by Interfacial Oxygen Octahedral Coupling." ACS Nano, **12 (4)**, 3681 (2018).

5 Sirui Zhang, Yinlian Zhu, Yunlong Tang, Ying Liu, Shuang Li, Mengjiao Han, Jinyuan Ma, B. Wu, Z. Chen, S. Saremi, and X. Ma, "Giant Polarization Sustainability in Ultrathin Ferroelectric Films Stabilized by Charge Transfer." Advanced Materials **29**, no. 46 (2017).

6 Shumin He, Guolei Liu, Yinlian Zhu, Xiuliang Ma, Jirong Sun, Shishou Kang, Shishen Yan, Yanxue Chen, Liangmo Mei, and Jun Jiao. "Impact of interfacial effects on ferroelectric resistance switching of Au/BiFeO 3/Nb: SrTiO 3 (100) Schottky junctions." RSC Advances **7**, no. 37, 22715 (2017)

7 D. Yadlovker, S. Berger. Uniform orientation and size of ferroelectric domains. Phys. Rev. B. **71**, 184112-1-6 (2005).

8 D. Yadlovker, S. Berger. Reversible electric field induced nonferroelectric to ferroelectric phase transition in single crystal nanorods of potassium nitrate. Appl. Phys. Lett. **91**, 173104 (2007).

9 D. Yadlovker, S. Berger. Nucleation and growth of single crystals with uniform crystallographic orientation inside alumina nanopores. J. Appl. Phys. **101**, 034304 (2007).

10. M. H. Frey, D. A. Payne. Grain-size effect on structure and phase transformations for barium titanate. Phys. Rev. B **54**, 3158- 3168 (1996).

11. Z. Zhao, V. Buscaglia, M. Viviani, M.T. Buscaglia, L. Mitoseriu, A. Testino, M. Nygren, M. Johnsson, P. Nanni. Grain-size effects on the ferroelectric behavior of dense nanocrystalline $BaTiO_3$ ceramics. Phys. Rev. B **70**, 024107-1-8 (2004).

12 A.V.Drobnich, A.A.Molnar, A.V.Gomonnai, Yu.M.Vysochanskii, I.P.Prits. The effect of size factor on the phase transition in Sn2P2S6 crystals: experimental data and simulation in ANNNI model. Condensed Matter Physics, **6**, 205 (2003)

13. E. Erdem, H.-Ch. Semmelhack, R. Bottcher, H. Rumpf, J. Banys, A.Matthes, H.-J. Glasel, D. Hirsch, E. Hartmann. Study of the tetragonal-to-cubic phase transition in $PbTiO_3$ nanopowders. J. Phys.: Condens. Matter **18**, 3861–3874 (2006).

14 T. Yu, Z. X. Shen, W. S. Toh, J. M. Xue, and J. Wang. Size effect on the ferroelectric phase transition in $SrBi_2Ta_2O_9$ nanoparticles J. Appl. Phys., **94**, 618, (2003)

15 I.S. Golovina, V.P. Bryksa, V.V. Strelchuk, I.N. Geifman, A.A. Andriiko. Size effects in the temperatures of phase transitions in $KNbO_3$ nanopowder. J. Appl. Phys**. 113**, 144103 (2013).

16. I.S. Golovina, V.P. Bryksa, V.V. Strelchuk, I.N. Geifman. Phase transitions in the nanopowders $KTa_{0.5}Nb_{0.5}O_3$ studied by Raman spectroscopy. Functional Materials. **20**, 75-80 (2013).

17 I.S. Golovina, B.D. Shanina, S.P. Kolesnik, I. N. Geifman, A. A. Andriiko. Magnetic properties of nanocrystalline $KNbO_3$. J. Appl. Phys. **114**, 174106 (2013).

18 L. M.Lopatina and J. Selinger. Theory of Ferroelectric Nanoparticles in Nematic Liquid Crystals. Physical Review Letters 102(19), 197802 (2009) doi: 10.1103/PhysRevLett.102.197802

19 Y. Garbovskiy and A. Glushchenko. Ferroelectric Nanoparticles in Liquid Crystals: Recent Progress and Current Challenges. Review. Nanomaterials, **7**, 361; (2017) doi:10.3390/nano7110361

20 P. Perriat, J. C. Niepce, G. Caboche. Thermodynamic considerations of the grain size dependency of material properties: a new approach to explain the variation of the dielectric permittivity of BaTiO3 with grain size. Journal of Thermal Analysis and Calorimetry **41**, 635-649 (1994).

21 H. Huang, C. Q. Sun, P. Hing. Surface bond contraction and its effect on the nanometric sized lead zirconate titanate. J. Phys.: Condens. Matter **12,** L127–L132 (2000).

22 H. Huang, C. Q. Sun, Z. Tianshu, P. Hing. Grain-size effect on ferroelectric $Pb(Zr_{1-x}Ti_x)O_3$ solid solutions induced by surface bond contraction. Phys. Rev. **B 63**, 184112 (2001).

23 M.D. Glinchuk, E.A. Eliseev, A.N. Morozovska, R. Blinc. Giant magnetoelectric effect induced by intrinsic surface stress in ferroic nanorods. Phys. Rev. **B 77**, 024106 (2008).

24 M.D. Glinchuk, E.A. Eliseev, A.N. Morozovska. Superparaelectric phase in the ensemble of noninteracting ferroelectric nanoparticles. Phys. Rev. **B. 78**, 134107 (2008).

25 M. Wenhui. Surface tension and Curie temperature in ferroelectric nanowires and nanodots. Appl. Phys. A **96**, 915–920 (2009).

26 V. V. Khist, E. A. Eliseev, M. D. Glinchuk, D. V. Karpinsky, M. V. Silibin, and A. N. Morozovska. Size Effects of Ferroelectric and Magnetoelectric Properties of Semi-ellipsoidal Bismuth Ferrite Nanoparticles. Journal of Alloys and Compounds, **714**, 15, 303–310 (2017)

27 J. Wang, A. K. Tagantsev, N. Setter. Size effect in ferroelectrics: Competition between geometrical and crystalline symmetries. Phys. Rev. B **83**, 014104 (2011)

28. A. N. Morozovska, E. A. Eliseev, M.D. Glinchuk. Ferroelectricity enhancement in confined nanorods: Direct variational method. Phys. Rev. B **73**, 214106 (2006).

29. A. N. Morozovska, M. D. Glinchuk, E.A. Eliseev. Phase transitions induced by confinement of ferroic nanoparticles. Phys. Rev. B **76**, 014102 (2007).

30 A.N. Morozovska, I.S. Golovina, S.V. Lemishko, A.A. Andriiko, S.A. Khainakov, E.A. Eliseev. Effect of Vegard strains on the extrinsic size effects in ferroelectric nanoparticles. Physical Review B **90**, 214103 (2014)

31 A. N. Morozovska, E. A. Eliseev, M. D. Glinchuk, O. M. Fesenko, V. V. Shvartsman, Venkatraman Gopalan, M. V. Silibin, D. V. Karpinsky. Rotomagnetic coupling in fine-grained multiferroic BiFeO3: theory and experiment. Phys.Rev. **B 97**,134115 (2018)

32 E.A. Eliseev, A.N. Morozovska, M.D. Glinchuk, R. Blinc. Spontaneous flexoelectric/flexomagnetic effect in nanoferroics. Phys. Rev. B. **79**, 165433-1-10, (2009).

33 E. A. Eliseev, A. V. Semchenko, Y. M. Fomichov, M. D. Glinchuk, V. V. Sidsky, V. V. Kolos, Yu M. Pleskachevsky, M. V. Silibin, N. V. Morozovsky, and A. N. Morozovska. "Surface and finite size effects impact on the phase diagrams, polar, and dielectric properties of (Sr, Bi) $Ta_2O_9$ ferroelectric nanoparticles." *Journal of Applied Physics* 119, no. 20: 204104 (2016).

34 E. A. Eliseev, V.V. Khist, M. V. Silibin, Y. M. Fomichov, G. S. Svechnikov, D. V. Karpinsky, V. V. Shvartsman, and A. N. Morozovska. Fixed Volume Effect on Polar Properties and Phase Diagrams of Ferroelectric Semi-ellipsoidal Nanoparticles. Accepted to EJPB (2018)

35 C.-G. Duan, S.S. Jaswal, E.Y.Tsymbal "Predicted magnetoelectric effect in Fe/BaTiO 3 multilayers: ferroelectric control of magnetism." Phys. Rev. Lett. **97**, 047201-1-4. (2006).

36 G. Geneste, E. Bousquest, J. Junquera, and P. Ghosez, "Finite-size effects in BaTiO3 nanowires." *Appl. Phys. Lett.* **88**, 112906 (2006).

37 M. Q. Cai, Y. Zheng, B. Wang, and G. W. Yang, "Nanosize confinement induced enhancement of spontaneous polarization in a ferroelectric nanowire." Appl. Phys. Lett. **95**, 232901 (2009).

38 J. W. Hong, G. Catalan, D. N. Fang, Emilio Artacho, and J. F. Scott. "Topology of the polarization field in ferroelectric nanowires from first principles" Phys. Rev. B 81, 172101 (2010).

39 E. Bousquet, N. Spaldin, and Ph. Ghosez, "Strain-Induced Ferroelectricity in Simple Rocksalt Binary Oxides" Phys. Rev. Lett. 104, 037601 (2010).

40 C.L. Wang, S.R.P.Smith "Landau theory of the size-driven phase transition in ferroelectrics" J Phys: Condens Matter **7**, 7163-7169 (1995).

41 A.N. Morozovska, E.A. Eliseev, R. Blinc, M.D. Glinchuk. Analytical prediction of size-induced ferroelectricity in BaO nanowires under stress. Phys. Rev. B 81, 092101 (2010).

42 A. Sundaresan, R. Bhargavi, N. Rangarajan, U. Siddesh, C.N.R. Rao. "Ferromagnetism as a universal feature of nanoparticles of the otherwise nonmagnetic oxides" Phys Rev B **74**, 161306(R) (2006).

43. D.D. Fong, G. B. Stephenson, S.K. Streiffer, J.A. Eastman, O.Auciello, P.H. Fuoss, C. Thompson. "Ferroelectricity in ultrathin perovskite films" *Science* **304**, 1650 (2004).

44 A.K. Tagantsev, L. E. Cross, and J. Fousek. *Domains in ferroic crystals and thin films*. New York: Springer, 2010. ISBN 978-1-4419-1416-3, e-ISBN 978-1-4419-1417-0, DOI 10.1007/978-1-4419-1417-0

45 E. A. Eliseev, Y. M. Fomichov, S. V. Kalinin, Y. M. Vysochanskii, P. Maksymovich and A. N. Morozovska. Labyrinthine domains on the phase diagram of ferroelectric nanoparticles: a manifestation of gradient-driven topological phase transition (http://arxiv.org/abs/1801.03545)

46 A. K. Tagantsev and G. Gerra. "Interface-induced phenomena in polarization response of ferroelectric thin films." J. Appl. Phys. **100**, 051607 (2006).

47 J. Bardeen, "Surface states and rectification at a metal semi-conductor contact." Phys. Rev. **71**, 717 (1947).

48 V.M. Fridkin, *Ferroelectrics semiconductors*, Consultant Bureau, New-York and London (1980). p. 119.

49 M.A. Itskovsky, "Some peculiarities of phase transition in thin layer ferroelectric." Fiz. Tv. Tela **16**, 2065 (1974).

50 P.W.M. Blom, R.M. Wolf, J.F.M. Cillessen, and M.P.C.M. Krijn. "Ferroelectric Schottky diode." Phys. Rev. Lett. **73**, 2107 (1994).

51 A.N. Morozovska, E.A. Eliseev, S.V. Svechnikov, A.D. Krutov, V.Y. Shur, A.Y. Borisevich, P. Maksymovych, and S.V. Kalinin. "Finite size and intrinsic field effect on the polar-active properties of ferroelectric semiconductor heterostructures." Phys. Rev. B. **81**, 205308 (2010).

52 Y.A. Genenko, O. Hirsch, and P. Erhart, "Surface potential at a ferroelectric grain due to asymmetric screening of depolarization fields." J. Appl. Phys. **115**, 104102 (2014).

53 G.B. StepIehenson and M.J. Highland, Equilibrium and stability of polarization in ultrathin ferroelectric films with ionic surface compensation. Physical Review **B**, 84 (6), p.064107 (2011)

54 M. J.Highland, T. T. Fister, D. D. Fong, P. H. Fuoss, Carol Thompson, J. A. Eastman, S. K. Streiffer, and G. B. Stephenson. "Equilibrium polarization of ultrathin PbTiO3 with surface compensation controlled by oxygen partial pressure." Physical Review Letters,**107**, no. 18, 187602 (2011).

55 K.Y. Foo, and B. H. Hameed. "Insights into the modeling of adsorption isotherm systems." *Chemical Engineering Journal* **156**.1: 2-10 (2010).

56 A. N. Morozovska, E. A. Eliseev, Y. A. Genenko, I. S. Vorotiahin, M. V. Silibin, Ye Cao, Yunseok Kim, M. D. Glinchuk, and S. V. Kalinin. Flexocoupling impact on the size effects of piezo- response and conductance in mixed-type ferroelectrics-semiconductors under applied pressure. *Phys. Rev.* B 94, 174101 (2016)

57 I. S. Vorotiahin, E. A. Eliseev, Qian Li, S. V. Kalinin, Y. A. Genenko and A. N. Morozovska. Tuning the Polar States of Ferroelectric Films via Surface Charges and Flexoelectricity. Acta Materialia 137 (15), 85–92 (2017)

58 E. A. Eliseev, I. S. Vorotiahin, Y. M. Fomichov, M. D. Glinchuk, S. V. Kalinin, Y. A. Genenko, and A. N. Morozovska. Defect driven flexo-chemical coupling in thin ferroelectric films. Physical Review B, 97, 024102 (2018)

59 D.R. Tilley. *Finite-size effects on phase transitions in ferroelectrics. Ferroelectic Thin Films*. ed. C. Paz de Araujo, J.F. Scott and G.W. Teylor. (Amsterdam: Gordon and Breach, 1996).

60 R. Kretschmer and K.Binder, Surface effects on phase transition in ferroelectrics and dipolar magnets. Phys. Rev. B. 20, 1065 (1979).

61 C.-L. Jia, V. Nagarajan, J.-Q. He, L. Houben, T. Zhao, R. Ramesh, K. Urban, and R. Waser. Unit-cell scale mapping of ferroelectricity and tetragonality in epitaxial ultrathin ferroelectric films, Nature Mat. **6**, 64 (2007).

62 Yu.M. Vysochanskii, M.M. Mayor, V.M. Rizak, V. Yu. Slivka, and M.M. Khoma. The tricritical Lifshitz point on the phase diagram of Sn2P2 (SexS,-x)6 ferroelectrics Zh. Eksp. Teor. Fiz. **95**, 1355 (1989) [Sov. Phys. JETP **68**, 782 (1989)].

63 A. N. Morozovska, E. A. Eliseev, C. M. Scherbakov, and Y. M. Vysochanskii, The influence of elastic strain gradient on the upper limit of flexocoupling strength, spatially-modulated phases and soft phonon dispersion in ferroics. *Phys. Rev.* **B 94**, 174112 (2016)

64 E. A. Eliseev, A. N. Morozovska, S. V. Kalinin, Yulan Li, Jie Shen, M. D. Glinchuk, Long-Qing Chen, and Venkatraman Gopalan. "Surface effect on domain wall width in ferroelectrics." Journal of Applied Physics 106, no. 8: 084102 (2009).

65 V.Yu.Korda, S.V.Berezovsky, A.S.Molev, L.P.Korda, V.F.Klepikov. On importance of higher non-linear interactions in the theory of type II incommensurate systems. Physica **B**, 425, 31–33 (2013) 66 R. Yevych, V. Haborets, M. Medulych, A. Molnar, A. Kohutych, A. Dziaugys, Ju. Banys, and Yu. Vysochanskii. Valence fluctuations in Sn(Pb)2P2S6 ferroelectrics. Low Temperature Physics/Fizika Nizkikh Temperatur, **42**, 1477 (2016)

67 V. Ongun Özçelik1 and A. Nihat Berker. Blume-Emery-Griffiths spin glass and inverted tricritical points. Phys. Rev. E **78**, 031104 (2008)

68 R. Yevych, M. Medulych, Yu. Vysochanskii. Nonlinear dynamics of ferroelectrics with three-well local potential. Accepted to Condensed Matter Physics (2018)

69 I. Zamaraite, R. Yevych, A. Dziaugys, A. Molnar, Ju. Banys, and Yu. Vysochanskii, Double hysteresis loops in proper uniaxial ferroelectrics (http://arxiv.org/abs/1805.08824)

70 Kenji Ishikawa, Kazutoshi Yoshikawa, and Nagaya Okada. "Size effect on the ferroelectric phase transition in $PbTiO_3$ ultrafine particles." *Physical Review B* 37, no. 10: 5852 (1988).

71 S. W. H. Eijt, R. Currat, J. E. Lorenzo, P. Saint-Gregoire, B. Hennion, and Yu M. Vysochanskii. "Soft modes and phonon interactions in studied by neutron scattering." The European Physical Journal B-Condensed Matter and Complex Systems 5, no. 2: 169-178 (1998).

72 A. N. Morozovska, E. A. Eliseev, JianJun Wang, G. S. Svechnikov, Yu M. Vysochanskii, Venkatraman Gopalan, and Long-Qing Chen. "Phase diagram and domain splitting in thin ferroelectric films with incommensurate phase." *Physical Review B* 81, no. 19: 195437 (2010).

73 A. Artemev, B. Geddes, J. Slutsker, and A. Roytburd. "Thermodynamic analysis and phase field modeling of domain structures in bilayer ferroelectric thin films." Journal of Applied Physics 103, no. 7: 074104 (2008).

74. A. Hubert and R. Schafer, *Magnetic Domains: The Analysis of Magnetic Microstructures*, Springer, **1998**.

75 G. Catalan, H. Béa, S. Fusil, M. Bibes, Patrycja Paruch, A. Barthélémy, and J. F. Scott. "Fractal dimension and size scaling of domains in thin films of multiferroic $BiFeO_3$." *Physical Review Letters* 100, no. 2: 027602 (2008).

76. S. V. Kalinin, B. J. Rodriguez, J. D. Budai, S. Jesse, A. N. Morozovska, A.A. Bokov, and Z.G. Ye, "Direct evidence of mesoscopic dynamic heterogeneities at the surfaces of ergodic ferroelectric relaxors." *Phys. Rev. B 81*, 064107 (2010).

77 S.V. Kalinin, A.N. Morozovska, L. Q. Chen, B. J. Rodriguez. Local polarization dynamics in ferroelectric materials. Rep. Prog. Phys. **73**, 056502-1-67 (2010).

78 A. Kholkin, A.Morozovska, D. Kiselev, I. Bdikin, B. Rodriguez, Pingping Wu, A. Bokov, Zuo-Guang Ye, Brahim Dkhil, Long-Qing Chen, M. Kosec, Sergei V. Kalinin. Surface Domain Structures and Mesoscopic Phase Transition in Relaxor Ferroelectrics. Advanced Functional Materials 21, No 11, 1977–1987 (2011)

79. V. V. Shvartsman and A. L. Kholkin, "Domain structure of 0.8 Pb (Mg 1/3 Nb 2/3) O 3− 0.2 PbTiO 3 studied by piezoresponse force microscopy." *Physical Review B* 69, 014102 (2004).

80. K. S. Wong, J. Y. Dai, X. Y. Zhao, and H. S. Luo. "Time-and temperature-dependent domain evolutions in poled (111)-cut (Pb (Mg 1∕3 Nb 2∕3) O 3) 0.7 (Pb Ti O 3) 0.3 single crystal." *Applied physics letters* 90, 162907 (2007).

81 A. Gruverman, O. Auciello, H. Tokumoto. Imaging and control of domain structures in ferroelectric thin films via scanning force microscopy. Annu. Rev. Mater. Sci. **28**, 101 (1998).

82 A. Gruverman, A. Kholkin. Nanoscale ferroelectrics: processing, characterization and future trends. Rep. Prog. Phys. **69**, 2443 (2006).

83 V. V. Shvartsman, A. L. Kholkin. Evolution of nanodomains in $0.9PbMg_{1/3}Nb_{2/3}O_3$-$0.1PbTiO_3$ single crystals. J. Appl. Phys. **101**, 064108 (2007).

84 A.N. Morozovska, E.A. Eliseev, S.L. Bravina, S.V. Kalinin. Resolution Function Theory in Piezoresponse Force Microscopy: Domain Wall Profile, Spatial Resolution, and Tip Calibration. Phys. Rev. **B 75**, 174109 (2007).

85 A.N. Morozovska, S.V. Svechnikov, E.A. Eliseev, S.V. Kalinin. Extrinsic Size Effect in Piezoresponse Force Microscopy of Thin Films. Phys. Rev. **B. 76**, 054123 (2007).

86 A. K.Yadav, C. T. Nelson, S. L. Hsu, Z. Hong, J. D. Clarkson, C. M. Schlepütz, A. R. Damodaran et al. "Observation of polar vortices in oxide superlattices." *Nature* 530, no. 7589: 198-201 (2016) doi:10.1038/nature16463

87 A. Gruverman, D. Wu, H-J Fan, I Vrejoiu, M Alexe, R J Harrison and J F Scott. "Vortex ferroelectric domains" J. Phys.: Condens. Matter 20, 342201, (2008) doi:10.1088/0953-8984/20/34/342201

88 N. Balke, B. Winchester, Wei Ren, Ying Hao Chu, A. N. Morozovska, E. A. Eliseev, M. Huijben, Rama K. Vasudevan, P. Maksymovych, J. Britson, S. Jesse, I. Kornev, Ramamoorthy Ramesh, L. Bellaiche, Long Qing Chen, and S. V. Kalinin. Enhanced electric conductivity at ferroelectric vortex cores in $BiFeO_3$. Nature Physics **8**, 81–88 (2012).

89 B. Winchester, N. Balke, X. X. Cheng, A. N. Morozovska, S. Kalinin, and L. Q. Chen. "Electroelastic fields in artificially created vortex cores in epitaxial $BiFeO_3$ thin films." Applied Physics Letters, **107**, (5), 052903 (2015).

90 Ivan I. Naumov, L. Bellaiche, and Huaxiang Fu. "Unusual phase transitions in ferroelectric nanodisks and nanorods." *Nature* 432, no. 7018: 737-740 (2004). doi: 10.1038/nature03107

91 A.N. Morozovska, E.A. Eliseev, Yulan Li, S.V. Svechnikov, P. Maksymovych, V.Y. Shur, Venkatraman Gopalan, Long-Qing Chen, and S.V. Kalinin, Thermodynamics of nanodomain formation and breakdown in Scanning Probe Microscopy : Landau-Ginzburg-Devonshire approach, Phys. Rev. B. 80, 214110 (2009).